\def\aSourceVector[#1]{{\rm \bf s}_{#1}}
\def\anObservationVector[#1]{{{\rm \bf x}_{#1}}}

\def\anElementOfaVectorInNullSpace[#1]{d_{#1}}
\def\aDictionaryColumn[#1]{{\rm \bf a}_{#1}}

\def\norm[#1]{\lVert{#1} \rVert}

\def\nullSpace[#1]{ \mathcal{N}(#1)}

\def\coherence[#1]{\mu(#1)}

\def\setR[#1][#2]{R_{\it #1}^{#2}}
\def\spark[#1]{{\rm spark}(#1)}
\def\krank[#1]{{\rm K\textrm{-}rank}(#1)}
\def\probability[#1]{P\left(#1\right)}
\def\absOperator[#1]{
\left|{#1}\right|
}
\def\cardinalityOperator[#1]{
\left\vert{#1}\right\vert
}
\def\arg\max\limits_{\theta}
\def\probability{\mathbb{P}}

\def\laplace[#1][#2]{{\mathscr{L}}_{#1}\left(#2\right)}
\def\bracket[#1]{\left[{#1}\right]}

\def\parantheses[#1]{\left({#1}\right)}

\documentclass[journal]{IEEEtran}
\IEEEoverridecommandlockouts

\pdfoutput=1
\usepackage{hyperref}
\pdftrailerid{}
\usepackage{amsthm}
\usepackage{amsmath}
\usepackage{amssymb}
\usepackage{booktabs}
\usepackage{siunitx}
\usepackage{cite}
\usepackage{graphics, framed}
\usepackage{graphicx}
\usepackage{epsfig}
\usepackage{epstopdf}
\usepackage{subfigure}
\usepackage{url}
\usepackage{multimedia}
\usepackage{paralist}
\usepackage[printonlyused]{acronym}
\usepackage{units}
\usepackage{balance}
\usepackage{multirow}

\usepackage{scalerel}
\usepackage{tikz}
\usetikzlibrary{svg.path}

\definecolor{orcidlogocol}{HTML}{A6CE39}
\tikzset{
  orcidlogo/.pic={
    \fill[orcidlogocol] svg{M256,128c0,70.7-57.3,128-128,128C57.3,256,0,198.7,0,128C0,57.3,57.3,0,128,0C198.7,0,256,57.3,256,128z};
    \fill[white] svg{M86.3,186.2H70.9V79.1h15.4v48.4V186.2z}
                 svg{M108.9,79.1h41.6c39.6,0,57,28.3,57,53.6c0,27.5-21.5,53.6-56.8,53.6h-41.8V79.1z M124.3,172.4h24.5c34.9,0,42.9-26.5,42.9-39.7c0-21.5-13.7-39.7-43.7-39.7h-23.7V172.4z}
                 svg{M88.7,56.8c0,5.5-4.5,10.1-10.1,10.1c-5.6,0-10.1-4.6-10.1-10.1c0-5.6,4.5-10.1,10.1-10.1C84.2,46.7,88.7,51.3,88.7,56.8z};
  }
}

\newcommand\orcidicon[1]{\href{https://orcid.org/#1}{\mbox{\scalerel*{
\begin{tikzpicture}[yscale=-1,transform shape]
\pic{orcidlogo};
\end{tikzpicture}
}{|}}}}

\usepackage{hyperref} %<- Load after everything else
\usepackage{tabularx,booktabs,ragged2e}
\def\trainingset{\mathbf{r}=\left(r_{1},r_{2},\cdots,r_{m}\right)}

\newcommand{\FGR}[1]{Fig.~\ref{#1}}

\newcommand{\SEC}[1]{Section~\ref{#1}}
\newcommand{\TAB}[1]{Table~\ref{#1}}
\newcommand{\EQ}[1]{Eq.~\eqref{#1}}

\usepackage{mleftright}

\newcommand{\lnb}[1]{%
\ln\mleft(#1\mright)%
}

\acrodef{5G}[5G]{5\textsuperscript{th}-Generation}
\acrodef{BW}[BW]{bandwidth}
\acrodef{CW}[CW]{continuous wave}
\acrodef{D2D}[D2D]{device-to-device}
\acrodef{M2M}[M2M]{machine-to-machine}
\acrodef{dB}[dB]{decibel}
\acrodef{dBi}[dBi]{decibel isotropic}
\acrodef{dBm}[dBm]{decibel over a milliwatt}
\acrodef{Gbps}[Gbps]{gigabit per second}
\acrodef{GHz}[GHz]{gigahertz}
\acrodef{Hz}[Hz]{hertz}
\acrodef{IF}[IF]{intermediate frequency}
\acrodef{IFFT}[IFFT]{inverse fast Fourier Transform}
\acrodef{KHz}[KHz]{kilohertz}
\acrodef{THz}[\unit{}{THz}]{terahertz}
\acrodef{LO}[LO]{local oscillator}
\acrodef{LOS}[LOS]{line-of-sight}
\acrodef{MHz}[MHz]{megahertz}
\acrodef{MILTAL}[M\.{I}LTAL]{Millimeter Wave and Terahertz Technologies Research Laboratories}
\acrodef{MIMO}[MIMO]{Multiple-input multiple-output}
\acrodef{SISO}[SISO]{single-input single-output}
\acrodef{mmWave}[mmWave]{millimeter wave}
\acrodef{NGWN}[NGWN]{next generation wireless network}
\acrodef{NLOS}[NLOS]{non line-of-sight}
\acrodef{OML}[OML]{Oleson Microwave Labs}
\acrodef{OLOS}[OLOS]{optical line-of-sight}
\acrodef{PNA}[PNA]{performance network analyzer}
\acrodef{QoS}[QoS]{quality of service}
\acrodef{RF}[RF]{radio frequency}
\acrodef{spar}[s-parameter]{Scattering parameter}
\acrodef{subThz}[sub-\unit{}{THz}]{sub-terahertz}
\acrodef{TUBITAK}[T\"{U}B\.{I}TAK]{Scientific and Technological Research Council of Turkey}
\acrodef{USB}[USB]{universal serial bus}
\acrodef{VNA}[VNA]{vector network analyzer}
\acrodef{AoA}[AoA]{angle of arrival}
\acrodef{MLE}[MLE]{maximum likelihood estimation}
\acrodef{FCC}[FCC]{Federal Communications Commission}

\acrodef{ADC}[ADC]{analog-to-digital converter}
\acrodef{ALMA}[ALMA]{Atacama large millimeter/sub-millimeter array}
\acrodef{AR}[AR]{augmented reality}
\acrodef{ASK}[ASK]{amplitude shift keying}
\acrodef{AWGN}[AWGN]{additive white Gaussian noise}

\acrodef{BAN}[BAN]{body area network} 
\acrodef{BER}[BER]{bit error rate}
\acrodef{BPSK}[BPSK]{binary phase shift keying}

\acrodef{CDF}[CDF]{cumulative distribution function}
\acrodef{CNT}[CNT]{carbon nano tube}
\acrodef{CTMA}[CTMA]{continuous-time moving-average}
\acrodef{CW}[CW]{continuous wave}

\acrodef{FEC}[FEC]{forward error correction}
\acrodef{FSK}[FSK]{frequency shift keying}
\acrodef{FSO}[FSO]{free space optical}
\acrodef{FSPL}[FSPL]{free space path loss}

\acrodef{GaN}[GaN]{gallium nitride}
\acrodef{Gbps}[Gbps]{gigabit per second}
\acrodef{GNR}[GNR]{graphene nano ribbon}
\acrodef{GPS}[GPS]{global positioning system}

\acrodef{HBT}[HBT]{heterojunction bipolar transistor}
\acrodef{HEMT}[HEMT]{high electron mobility transistor}
\acrodef{HITRAN}[HITRAN]{high-resolution transmission}

\acrodef{IF}[IF]{intermediate frequency}
\acrodef{InP}[InP]{indium phosphide}
\acrodef{I/Q}[I/Q]{in-phase and quadrature}
\acrodef{KS}[KS]{Kolmogorov-Smirnov}
\acrodef{LO}[LO]{local oscillator}
\acrodef{LOS}[LOS]{line-of-sight}

\acrodef{MAC}[MAC]{medium access control}
\acrodef{MGF}[MGF]{moment generating function}
\acrodef{MILTAL}[M\.{I}LTAL]{Millimeter Wave and Terahertz Technologies Research Laboratories}
\acrodef{MIMO}[MIMO]{Multi-input multi-output}
\acrodef{MLE}[MLE]{maximum likelihood estimation}

\acrodef{NLOS}[NLOS]{non-line-of-sight}
\acrodef{OFDM}[OFDM]{orthogonal frequency division multiplexing}
\acrodef{OML}[OML]{Oleson Microwave Labs}
\acrodef{OOK}[OOK]{on-off keying}

\acrodef{PCB}[PCB]{printed circuit board}
\acrodef{PDF}[PDF]{probability density function}
\acrodef{PNA}[PNA]{performance network analyzer}
\acrodef{PPM}[PPM]{pulse position modulation}
\acrodef{PSK}[PSK]{phase shift keying}
\acrodef{QAM}[QAM]{quadrature amplitude modulation}
\acrodef{QCL}[QCL]{quantum cascade laser}
\acrodef{QPSK}[QPSK]{quadrature phase shift keying}
\acrodef{RF}[RF]{radio frequency}

\acrodef{SiGe}[SiGe]{silicon germanium}
\acrodef{SNR}[SNR]{signal-to-noise ratio}
\acrodef{SPP}[SPP]{surface plasmon polariton}
\acrodef{SSA}[SSA]{Signal and Spectrum Analyzer}

\acrodef{Tbps}[Tbps]{terabits per second}
\acrodef{TCP}[TCP]{transmission control protocol}
\acrodef{TS-OOK}[TS-OOK]{time spread \ac{OOK}}
\acrodef{TUBITAK}[T\"{U}B\.{I}TAK]{Scientific and Technological Research Council of Turkey}

\acrodef{UHF}[UHF]{ultra high frequency}
\acrodef{USB}[USB]{universal serial bus}
\acrodef{VNA}[VNA]{vector network analyzer}
\acrodef{VR}[VR]{virtual reality}

\acrodef{WMRD}[WMRD]{Weighted relative mean difference}
\acrodef{WPAN}[WPAN]{wireless personal area network}
\acrodef{THz}[THz]{terahertz}
\acrodef{EM}[EM]{expectation-maximization}
\acrodef{KL}[KL]{Kullback-–Leibler}
\acrodef{ML}[ML]{maximum likelihood}
\acrodef{expE}[E]{expectation}
\acrodef{maxM}[M]{maximization}

\ifCLASSINFOpdf
\else
\fi

\begin{document}
\title{Modeling and Analysis of Short Distance Sub-Terahertz Communication Channel via Mixture of Gamma Distribution}
  \author{K{\"{u}}r{\c{s}}at Tekb{\i}y{\i}k,~\IEEEmembership{Graduate Student Member,~IEEE,} Ali R{\i}za Ekti,~\IEEEmembership{Senior~Member,~IEEE,}\\G{\"{u}}ne{\c{s}} Karabulut Kurt,~\IEEEmembership{Senior Member,~IEEE,} Ali G\"{o}r\c{c}in,~\IEEEmembership{Senior Member,~IEEE}\\Serhan Yarkan,~\IEEEmembership{Senior Member,~IEEE}   

 \thanks{K. Tekb{\i}y{\i}k and G.K. Kurt are with the Department of Electronics and Communications Engineering, {\.{I}}stanbul Technical University, {\.{I}}stanbul, Turkey.~~e-mail: \{tekbiyik, gkurt\}@itu.edu.tr}
 \thanks{K. Tekb{\i}y{\i}k,  A. G\"{o}r\c{c}in and A.R. Ekti are with the HISARLab at Informatics and Information Security Research Center (B{\.{I}}LGEM), T{\"{U}}B{\.{I}}TAK, Kocaeli, Turkey.~~e-mail: \{kursat.tekbiyik, ali.gorcin, aliriza.ekti\}@tubitak.gov.tr} 
\thanks{A.R. Ekti is with the Department of Electrical-Electronics Engineering, Bal{\i}kesir University, Bal{\i}kesir, Turkey.~~e-mail: arekti@balikesir.edu.tr}
\thanks{A. G\"{o}r\c{c}in is with the Department of Electronics and Communications Engineering, Y{{\i}}ld{{\i}}z Technical University, {\.{I}}stanbul, Turkey.~~e-mail: agorcin@yildiz.edu.tr}
\thanks{S. Yarkan is with the Department of Electrical and Electronics Engineering, {\.{I}}stanbul Commerce University, {\.{I}}stanbul, Turkey.~~e-mail: syarkan@ticaret.edu.tr}
}

\IEEEoverridecommandlockouts

\maketitle

\begin{abstract}

With the recent developments on opening the \ac{THz} spectrum for experimental purposes by the \acl{FCC}, transceivers operating in the range of $0.1$\ac{THz}-$10$\ac{THz}, which are known as \ac{THz} bands, will enable ultra-high throughput wireless communications. However, actual implementation of the high-speed and high reliability \ac{THz} band communication systems should start with providing extensive knowledge in regards to the propagation channel characteristics. Considering the huge bandwidth and the rapid changes in the characteristics of \ac{THz} wireless channels, ray tracing and one-shot statistical modeling are not adequate to define an accurate channel model. In this work, we propose Gamma mixture based channel modeling for the \ac{THz} band via the \ac{EM} algorithm. First, \ac{MLE} is applied to characterize the Gamma mixture model parameters, and then \ac{EM} algorithm is used to compute \acp{MLE} of the unknown parameters of the measurement data. The accuracy of the proposed model is investigated by using the \ac{WMRD} error metrics, \ac{KL}-divergence, and \ac{KS} test to show the difference between the proposed model and the actual \acp{PDF} that are obtained via the designed test environment. To efficiently evaluate the performance of the proposed method in more realistic scenarios, all the analysis is done by examining measurement data from a measurement campaign in the \unit{$240$}{GHz} to \unit{$300$}{GHz} frequency range, using a well-isolated anechoic chamber. According to \ac{WMRD} error metrics, \ac{KL}-divergence, and \ac{KS} test results, \acp{PDF} generated by the mixture of Gamma distributions fit to the actual histogram of the measurement data. It is shown that instead of taking pseudo-average characteristics of sub-bands in the wide band, using the mixture models allows for determining channel parameters more precisely.
\end{abstract}

\IEEEpeerreviewmaketitle
\acresetall

%%%%%%%%%%%%%%%%%%%%%%%%%%%%%%%% INTRODUCTION %%%%%%%%%%%%%%%%%%%%%%%%
\section{Introduction}\label{sec:intro}
%%%%%%%%%%%%%%%%%%%%%%%%%%%%%%%% INTRODUCTION %%%%%%%%%%%%%%%%%%%%%%%%

The ability of wireless communication technology to meet consumer needs requires that next generation wireless networks' data rates reach \ac{Tbps} levels at a higher link density \cite{cisco2017zettabyte, akyildiz2014terahertz}. Although \ac{FSO} and \ac{mmWave} communications are proposed for high data rates, the requirements of both systems, and especially a bandwidth of only $9$GHz around $60$GHz, are not expected to deliver \ac{Tbps} for mobile and personal communication systems \cite{8276837}. As there is no block wider than $10$GHz below $100$GHz \cite{tekbiyik2019terahertz}, the researchers push the frequency limits towards \ac{THz} band, which is in between $0.1$THz - $10$THz. Due to the flat frequency response and also the capabilities of the current state of the signal generators, most of the researches focus on the band between $200$GHz - $300$GHz.

To be able to fully discover the potential of a wireless communication system, the proper channel model must be used. Then, other parts of the system can be designed. Although the \ac{THz} band will provide a way to achieve \ac{Tbps} data rates, the \ac{THz} band differs from the currently used bands in the channel characteristics that change rapidly and sharply across the spectrum \cite{akyildiz2014terahertz, tekbiyik2019terahertz}. Therefore, all elements of the system should be re-examined and designed to develop a proper communication system. For example, the propagation channel is required to be analyzed on the aspects of materials in the medium, and the operating frequency. Wireless communication in wide band around $60$GHz requires a channel model considering characteristics of sub-bands which are windows such that propagation characteristics can be assumed to be static throughout the window. 

\subsection{Related Works}
In the studies on channel modeling, various approaches can be encountered for frequency, time and spatial analysis. It is worth saying that THz channels have many differences from lower frequency bands in terms of noise, propagation, and molecular absorption. Thus, channel modeling studies require to be reconsidered for THz band rather than employ the model proposed for lower bands. The wireless communication channel can be modeled by using deterministic or statistical methods. Although deterministic models like ray-tracing are most accurate if the detailed description of the given environment is properly and extensively fed into \cite{peng2016channel}, their performance can be hindered even in the presence of a slightest change in the propagation environment. All parameters of the propagation environment are required by these models. As a result, considering that even molecular changes affect the propagation characteristics of \ac{THz} waves, it can be said that the ray-tracing method might not adequately model THz channels. Another reason that makes this method complex and computationally cumbersome is the exponential increase in the complexity of the method as the size of the medium to be modeled increases. On the contrary, temporospatial characteristics of wireless channels of data centers are investigated  in \cite{peng2015stochastic, nie2017three}. Furthermore, multi-dimensional parameters of kiosk's wireless channels are modelled for each type of \ac{THz} rays in \cite{he2017stochastic}. The statistical approaches use the average of the environmental effects, unlike a deterministic model (e.g., ray-tracing). Some stochastic models have been recently proposed in~\cite{priebe2013stochastic,han2015multi,ekti2017statistical}. Our previous work \cite{tekbiyik2019statistical} proposes a two-slope path loss model for short-range \ac{THz} communication links. In \cite{ju20203}, the statistical channel parameters such as delay-spread, cluster delays and cluster powers are obtained throughout extensive measurements at $140$GHz. It should be noted that that ray-tracing and stochastic methods are not antipodes; but, they are complement of each other to describe a channel more accurate. For example, \cite{chen2021channel} proposes a hybrid channel model by combining statistical approaches and ray-tracing methods for sub-THz band.

Another important consideration in channel modeling is the careful selection of signal processing methods to be used for modeling the wide-band channel. To set an example, in \cite{khalid2016wideband}, the frequency sweeping method, which is not safe due to artifacts created when the post-processing of the smaller chunks of bandwidth, is employed to model the spectrum between \unit{$260$}{GHz} and \unit{$400$}{GHz}. Another problem in channel modeling is to make the assumption that the derived impulse response has a linear phase. This assumption implies that the impulse response is symmetrical to \ac{LOS} propagation delay. However, the real physical environments do not allow this phenomenon because it contradicts causality. Therefore, Kazuhiro \textit{et. al.} propose a causal channel model for \ac{THz} band \cite{tsujimura2018causal}.

\ac{MIMO} can provide coverage improvements in addition to capacity enhancements for \ac{THz} communications; thus, channel models for $2\times2$ \ac{MIMO} systems are investigated in \cite{xu2014design,khalid2016experimental}. The results indicate that \ac{MIMO} systems can achieve high data rates. In~\cite{cheng2020terahertz}, Doppler shift caused by airflow turbulance in data-center is measured for band between $300$GHz and $320$GHz. Besides Doppler shift measurements, it presents that channel amplitude gains for $4\times4$ \ac{MIMO} follow m-Nakagami distribution. Furthermore, the cluster shadowing gain in a data-center is Gaussian distributed for that band~\cite{cheng2020thz}. Also, by using graphene-based \ac{MIMO} system, the spectral efficiency can be enhanced. Massive \ac{MIMO} systems benefits most from the ultra small antenna sizes at \ac{THz} frequencies; therefore, massive \ac{MIMO} antenna structures for these bands are researched in \cite{akyildiz2016realizing,zakrajsek2017design,han2018ultra}. These inquiries show that the capabilities of the \ac{THz} communication can be advanced by utilizing nano antenna structures and massive \ac{MIMO} systems. Besides these works, some studies focus on the application specific aspects of these bands; in \cite{priebe2011channel}, indoor channel measurements are conducted for \unit{$300$}{GHz}. Also, in \cite{hudlivcka2017characterization}, the behavior of the digital communication schemes are analyzed for the same band.

Studies up to this point assume that the \ac{THz} band of interest has a single statistical distribution. In this case, it can be concluded that channel modeling with a single \ac{PDF} is not sufficient, considering the presence of windows that behave differently in the \ac{THz} band due to the effect of molecules in the medium. The \ac{THz} band contains changes across the spectrum, so it may not be sufficient to express this extremely wide-band with a single statistical model. For example, suppose that the three sub-bands behave differently from each other as demonstrated in \FGR{fig:mixture_spectrum}. Hence, the use of mixtures to add the characteristics of each sub-band into the model provides better convergence to the actual histogram. It is worth noting that an arbitrary \ac{PDF} can be modeled by utilizing Gamma mixtures~\cite{wiper2001mixtures, da2015algorithm}. Therefore, in this study, we investigate the channel modeling with Gamma mixtures for short-range sub-THz channels.

Although mixture models have been used in many different fields, in this study, we are confined to mentioning only the studies on wireless communication channel models. In \cite{al2017mixture}, it is stated that mixture Gamma is able to model $\alpha-\kappa-\mu$ shadowed fading channels, even though they consist of intractable statistical properties. \cite{alhussein2015generalized} and \cite{selim2015modeling} employ the mixture of Gaussian distributions to construct a generalized shadowing model, by adopting \ac{EM} algorithm to find the mixture parameters. The error probability and ergodic capacity can be analyzed by using Gamma mixtures for diversity reception schemes over generalized-$K$ fading channels \cite{jung2014capacity}. Moreover, the physical layer security analysis can be performed by utilizing mixture models in generalized-$K$ fading channels \cite{lei2015performance}. \cite{atapattu2011mixture}, which is one of the most important studies in this research area, proposes a mixture of Gamma distributions for the \ac{SNR} of fading channels; thereby, it allows to derive the average channel capacity, the outage probability, and the symbol error rate.

\begin{figure*}[!t]
    \centering
    \includegraphics[width=\linewidth]{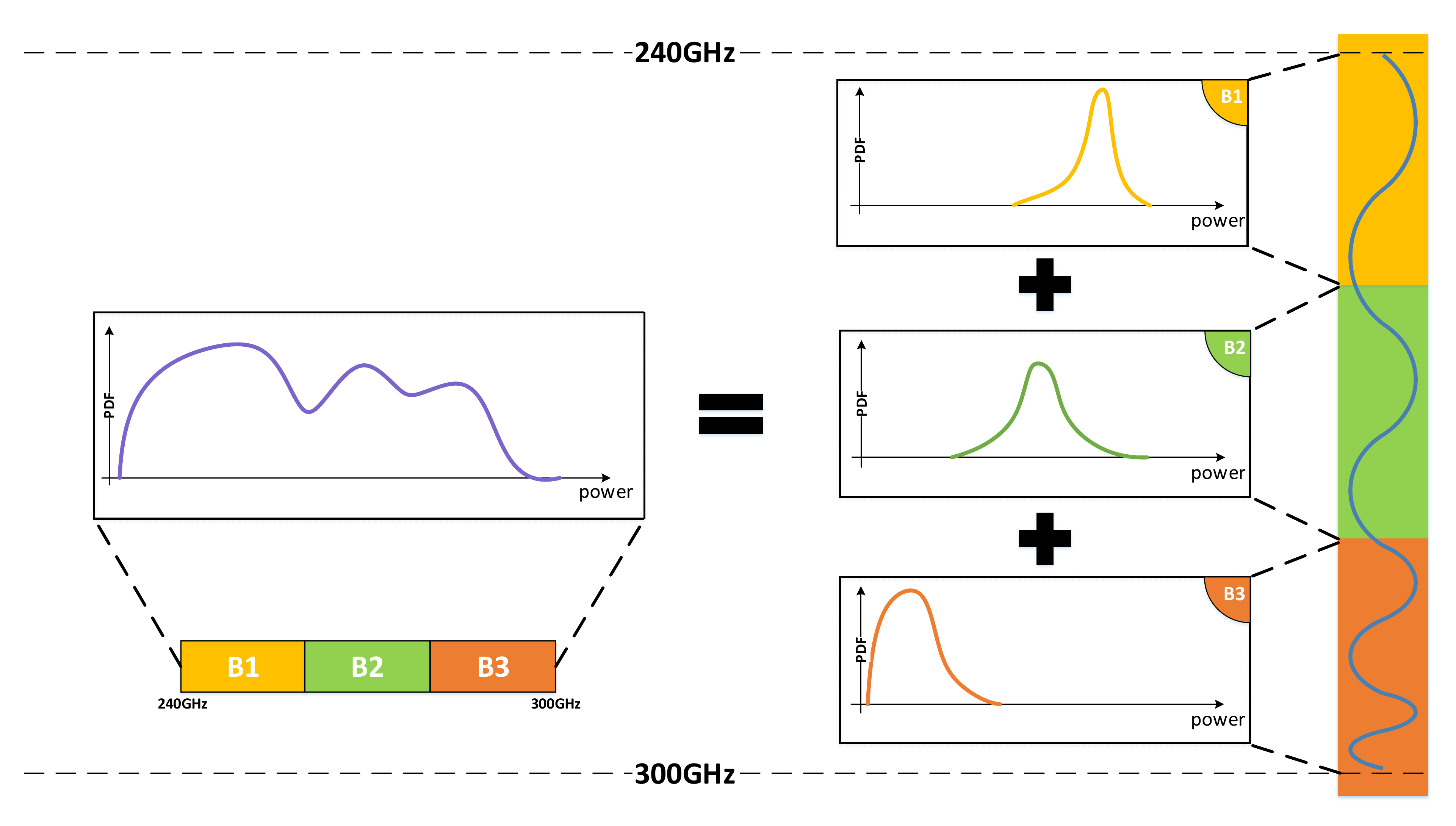}
    \caption{Instead of using a single distribution to model the received power characteristics, mixture model is able to give information about each sub-band characteristics. It can be said that the wide-band signal covers the characteristics of each sub-band signal propagates in the bands B1, B2, and B3.}
    \label{fig:mixture_spectrum}
\end{figure*}
%%%%%%%%%%%%%%%%%%%%%%%%%%%%%%%% Contributions %%%%%%%%%%%%%%%%%%%%%%%%
\subsection{Contributions}
In this study, we utilize mixture models to investigate channel models for sub-\ac{THz} band between \unit{$240$}{GHz} and \unit{$300$}{GHz}. With inspiration from the studies such as \cite{cellini2005novel, buyukccorak2015lognormal}, which adopt the mixture models to characterize the wireless propagation channel, we propose mixture models which are suitable for the nature of the \ac{THz} band to model the distribution of the received power for the sub-\ac{THz} band between \unit{$240$}{GHz} and \unit{$300$}{GHz}. Based on the intrinsic propagation characteristics of the wide-band THz communication channel, a channel model based on a single distribution can not provide an adequate representation. As the \ac{THz} band allows for very broadband communication and there is a significant change \cite{tekbiyik2019statistical} in channel characteristics throughout this wide band. The contributions of this study can be categorized under three main points:
\begin{itemize}
    \item For the \ac{THz} band, measurement based channel model study is performed. Using measurement data, it is shown that Gamma mixtures can be used effectively in channel modeling for \ac{THz} band. Thus, the characteristics of the channel can be expressed in a realistic manner.
    \item \ac{WMRD}, \ac{KS} and \ac{KL}-divergence approaches are studied to investigate how well Gamma mixture models fit into measurement data.
    \item Moreover, considering that the measurement data used in this study is a very valuable source of information and the necessity of making serious investments to reach such data, it is offered as a public dataset \cite{tekbiyik2019thz}. We believe that the sharing of this measurement data will foster new studies.
\end{itemize}
%%%%%%%%%%%%%%%%%%%%%%%%%%%%%%%% Contributions %%%%%%%%%%%%%%%%%%%%%%%%

%%%%%%%%%%%%%%%%%%%%%%%%%%%%%%%% Org. of the Paper %%%%%%%%%%%%%%%%%%%%%%%%
\subsection{Organization of the Paper}

The rest of this manuscript is organized as follows. \SEC{sec:background} details the signal model and gives mathematical preliminaries. The measurement setup is introduced in \SEC{sec:measurement_setup}. In \SEC{sec:gamma_mix_thz}, Gamma mixture modeling results are given and discussed. Finally, \SEC{sec:conclusion} concludes the study.
%%%%%%%%%%%%%%%%%%%%%%%%%%%%%%%% Org. of the Paper %%%%%%%%%%%%%%%%%%%%%%%%

%%%%%%%%%%%%%%%%%%%%%%%%%%%%%%%% BACKGROUND %%%%%%%%%%%%%%%%%%%%%%%%
\section{Background}\label{sec:background}
%%%%%%%%%%%%%%%%%%%%%%%%%%%%%%%% BACKGROUND %%%%%%%%%%%%%%%%%%%%%%%%

%%%%%%%%%%%%%%%%%%%%%%%%%%%%%%%% Signal Model %%%%%%%%%%%%%%%%%%%%%%%%
\subsection{Signal Model}\label{sec:signal_model}
The received signal is represented as:
\begin{align}
r(t) = \mathrm{Re}\big\{[x_I(t) + jx_Q(t)]e^{j2\pi f_c t}\big\},
\end{align}
where $j$ denotes the unit imaginary number and $\mathrm{Re}\{\cdot\}$ is the real part of the complex number. $x_I(t)$ and $x_Q(t)$ are \ac{I/Q} parts of the complex baseband signal. $f_c$ stands for the carrier frequency of the signal. 

The multipath channel at passband with different delays and attenuation levels can be given as:
\begin{align}
h(t) = \sum_{l=0}^{L-1} a_l \delta(t-t_l),
\label{eq:channel_passband}
\end{align}
where $L$ is the number of multipath components. $a_l$ and $t_l$ denote the attenuation and delay factors for the $l$th path, respectively. The complex baseband representation of \EQ{eq:channel_passband} is
\begin{align}
h(t) = \sum_{l=0}^{L-1} a_l \delta(t-t_l)e^{-j2\pi f_c t_l}.
\label{eq:channel_baseband}
\end{align}

If the channel consists of only \ac{LOS} component, $L$ in \EQ{eq:channel_baseband} is equal to 1. Then, \ac{LOS} channel is given as:
\begin{align}
h(t) = a_0 \delta(t-t_0)e^{-j2\pi f_c t_0},
\label{eq:channel_baseband_loss}
\end{align}
where $a_0$ and $2\pi f_c t_0$ denote amplitude and phase of channel, respectively. $t_0$ is propagation delay given with
\begin{align}
t_0 = \frac{d}{c},
\end{align}
where $d$ is the distance between transmitter and receiver and $c$ is the speed of light.

Anechoic chambers, as used in our measurements, do not allow \ac{NLOS} propagation. The losses are limited to antenna misalignment, imperfections created by hardware, and path loss. Thus, the signal model can be reduced to a direct path which is comprised of distant dependent path loss and antenna misalignment. The contribution of path loss to the channel amplitude $a_0$ is given as:
\begin{align}
P_{RX} = P_{TX} - 10n\mathrm{log}(d) + M.
\end{align}
The received power $P_{RX}$ including antenna gain considering misalignment, $M$, is calculated as the difference between transmitted power $P_{TX}$ and path loss with exponent $n$.
%%%%%%%%%%%%%%%%%%%%%%%%%%%%%%%% Signal Model %%%%%%%%%%%%%%%%%%%%%%%%

%%%%%%%%%%%%%%%%%%%%%%%%%%%%%%%% Gamma Distribution %%%%%%%%%%%%%%%%%%%%%%%%
\subsection{Gamma Distribution}\label{sec:gamma_why}

The Gamma function, $\Gamma(a)$, is defined as \cite{ramachandran2014mathematical}:
\begin{align}
\Gamma(a) = \int_{0}^{\infty}e^{-x}x^{a-1}dx, \; a>0.
\end{align}
By using integration by parts, $\Gamma(a)=(a-1)!$ when $a$ is a positive integer. Consider the random variable $G$ which is a mixture of $m$ Gamma distributions and defined as:

\begin{equation}
\label{eq:gm_mix_general}
f_G(x)=\sum_{l=1}^{m}\rho_l f_l(x;\alpha_l,\beta_l),~l=1,2,\ldots,m,~~x>0,~~\rho_l>0
\end{equation}

\noindent where $f_l(x;\alpha_l,\beta_l)=\frac{1}{\beta_l^{\alpha_l}\Gamma(\alpha_l)}x^{\alpha_l-1}e^{-x/\beta_l}$; $\alpha_l>0$ and $\beta_l>0$ are the shape and scale parameters of the $l$th component of the mixture distribution; $\rho_l$ denotes mixture proportions or weights that satisfy the conditions \begin{inparaenum}[(a)]\item{$0<\rho_l<1,~\forall l=1,2,\ldots,m$} and \item{$\sum_{l=1}^{m}\rho_l=1$}\end{inparaenum}. Please note that, $m$ denotes the number of components in the mixture. The main reasons for using a mixture of Gamma distributions in the paper are: \begin{inparaenum}[(i)]\item{the tractability of its \ac{CDF} and \ac{MGF}}, \item{giving an approximation for small-scale fading channels \cite{atapattu2011mixture}}, and \item{high accuracy by properly adjusting parameters}\end{inparaenum}.
%%%%%%%%%%%%%%%%%%%%%%%%%%%%%%%% Gamma Distribution %%%%%%%%%%%%%%%%%%%%%%%%
%%%%%%%%%%%%%%%%%%%%%%%%%%%%%%%% Max. Like. Est. %%%%%%%%%%%%%%%%%%%%%%%%
\subsection{Maximum Likelihood Estimation}

The \ac{MLE} technique is provided to obtain the parameters of the gamma mixture from the actual channel \ac{PDF}. Let assume that $X_{1},\cdots,X_{n}$ are random variables with Gamma distribution (with unknown parameters $\alpha>0$ and $\beta>0$). The likelihood function is given as:
\begin{align}
\label{eq:likelihood_gamma}
L(x;\alpha, \beta) &= \prod_{i=1}^{n} \frac{x_i^{\alpha-1}e^{-\frac{x_i}{\beta}}}{\Gamma\left(\alpha\right)\beta^{\alpha}}\nonumber\\&=\left\{\prod_{i=1}^{n}x_i\right\}^{-1}\left\{\prod_{i=1}^{n}x_i\right\}^ae^{-\frac{\sum_{i=1}^{n} x_i}{\beta}}\beta^{-n\alpha}\Gamma^{-n}\left(\alpha\right)\nonumber\\&\hspace*{-1.4cm}\text{The uninformative factor, $\left\{\prod_{i=1}^{n}x_i\right\}^{-1}$, is discarded}\nonumber\\&=\left\{\prod_{i=1}^{n}x_i\right\}^ae^{-\frac{\sum_{i=1}^{n} x_i}{\beta}}\beta^{-n\alpha}\Gamma^{-n}\left(\alpha\right)\nonumber\\&=\beta^{-n\alpha}\Gamma^{-n}\left(\alpha\right)\left\{\prod_{i=1}^{n}x_i\right\}^ae^{-\frac{\sum_{i=1}^{n} x_i}{\beta}}
\end{align}
The corresponding log likelihood function of \EQ{eq:likelihood_gamma} leads to:
\begin{align}
\label{eq:log_likelihood_gamma}
\lnb{L}&=-n\alpha\lnb{\beta}-n\lnb{\Gamma(a)}+\alpha\sum_{i=1}^{n}\lnb{x_i}-\sum_{i=1}^{n}\frac{x_i}{\beta}.
\end{align}
Maximum likelihood estimates can be found for $\alpha$ and $\beta$ by taking partial derivatives of \EQ{eq:log_likelihood_gamma} with respect to $\alpha$ and $\beta$, then we obtain:
\begin{align}
\label{eq_alpha_beta_mle_calculation}
\frac{\partial \lnb{L}}{\partial \alpha}&=-n\lnb{\beta}-n\frac{\partial \lnb{\Gamma(a)}}{\partial \alpha}+\sum_{i=1}^{n}\lnb{x_i}\nonumber\\
\frac{\partial \lnb{L}}{\partial \beta}&=-n\alpha\frac{1}{\beta}+\sum_{i=1}^{n}\frac{x_i}{\beta^2}
\end{align}
Because of the diGamma and logarithm functions in \EQ{eq_alpha_beta_mle_calculation}, a closed-form solution could not be provided \cite{ramachandran2014mathematical}. Numerical methods such as Newton-Raphson can be applied to find the values for $\alpha$ and $\beta$ which is not the scope of this study.
%%%%%%%%%%%%%%%%%%%%%%%%%%%%%%%% Max. Like. Est. %%%%%%%%%%%%%%%%%%%%%%%%

%%%%%%%%%%%%%%%%%%%%%%%%%%%%%%%% Exp. Maximization %%%%%%%%%%%%%%%%%%%%%%%%
\subsection{Expectation Maximization}
We have a training set $\trainingset$ consisting of $m$ independent observations captured by considering each measurement data at different transmitter-receiver separation distances such as $d$=\unit{$20$}{cm}, \unit{$30$}{cm}, \unit{$40$}{cm}, \unit{$60$}{cm}, \unit{$80$}{cm}. Our goal is to fit the Gamma distribution parameters by utilizing the \ac{EM} algorithm. \ac{EM} algorithm, which is a machine learning technique \cite{murphy2012machine}, provides a simplification to \ac{MLE} problems, which are mostly seen in mixture models \cite{buyukccorak2015lognormal}. The \ac{EM} algorithm consists of two steps, namely, the \ac{expE}-step and the \ac{maxM}-step. The reader is referred to \cite{em_dempster} for more detailed explanations about the \ac{EM} algorithm.

The \ac{EM} algorithm requires number of mixtures as a priori. Initially, the parameters are randomly chosen for the mixture model parameters $\theta_{1:M} = (\theta_{1},\cdots,\theta_{M})$. Then, the parameters are updated in each iteration until the convergence criteria hold. E-step calculates membership coefficients for all data point ($i=1,\cdots,L$) and mixture components ($k=1,\cdots,M$) by utilizing the current parameters $\theta_{1:M}$ \cite{buyukccorak2015lognormal,bilmes1998gentle}
\begin{align}
\phi_{ik} = \frac{\pi_{k}p_{k}(x_{i}|\theta_{k})}{\sum_{k=1}^{M}\pi_{k}p_{k}(x_{i}|\theta_{k})},
\end{align}
where $x_i$ is the data in the $k$th mixture; $\pi_{k}$ denotes the mixing proportion. It is obvious that $\sum_{k=1}^{M}\phi_{ik}=1$. Then, the parameter values and the mixing proportions for each mixture components are updated to maximize the likelihood probability in the M-step. In the M-step, the membership coefficients calculated in E-step are used to find parameters and mixing proportions as:
\begin{align}
\pi_{k}^{new}&=\frac{\sum_{i=1}^{L}\phi_{ik}}{L} \nonumber \\
\mathbb{E}[X_k]^{new} &= \frac{\sum_{i=1}^ {L}\phi_{ik}x_i}{\sum_{i=1}^{L}\phi_{ik}} = \alpha \beta \nonumber \\
\mathrm{Var}[X_k]^{new} &= \frac{\sum_{i=1}^ {L}\phi_{ik}(x_i-\mathbb{E}[X_k]^{new})^2}{\sum_{i=1}^{L}\phi_{ik}} = \alpha \beta^2.
\label{eq:parameters}
\end{align}
The parameters ($\alpha, \beta$) for each Gamma mixture can be found by using \EQ{eq:parameters}.
%%%%%%%%%%%%%%%%%%%%%%%%%%%%%%%% Exp. Maximization %%%%%%%%%%%%%%%%%%%%%%%%
%%%%%%%%%%%%%%%%%%%%%%%%%%%%%%%% Error Metrics %%%%%%%%%%%%%%%%%%%%%%%%
\subsection{Error Metrics}
In this subsection, we provide an overview of the possible error metrics to determine the goodness-of-fit for the proposed model. 
\subsubsection{Weighted Mean Relative Difference}
The proposed models are quantified by using \ac{WMRD}, which gives a measurement for the difference between the model and actual \acp{PDF}. It is defined as \cite{buyukccorak2015lognormal}:
\begin{align}
\mathrm{WMRD} = \frac{\sum_{\rho}|y_\rho-\hat{y}_\rho|}{\sum_{\rho}(y_\rho+\hat{y}_\rho)\times0.5},
\end{align}
where $\rho$ represents the received power and $y_\rho$ is the number of $\rho$ value observations in the received power set. As well as, $\hat{y}_\rho$ is related to the estimated model. 

\subsubsection{Kolmogorov-Smirnov Test}
\ac{KS} test is a non-parametric goodness-of-fit test, namely it does not make an assumption of any distribution. In addition to vector norm based error technique, the \ac{KS} test is employed as goodness-of-fit test with the confidence level $p=0.05$ to compare the actual \ac{PDF} with the estimated mixture models.

\subsubsection{Kullback–-Leibler Divergence}
\ac{KL} distance or divergence is interpreted as the distance between the actual probability distribution, $P_{act}$ and the estimated probability distribution, $P_{est}$. Let $P_{act} = \left\{p_1, p_2, \cdots, p_n \right\}$ and $P_{est} = \left\{q_1, q_2, \cdots, q_n \right\}$, then \ac{KL}-divergence is defined as
\begin{align}
    D_{\mathrm{KL}}(P_{act} \| P_{est})=-\sum_{x \in \mathcal{X}} P_{act}(x) \log \left(\frac{P_{est}(x)}{P_{act}(x)}\right).
\end{align}
In this paper, \ac{KL}-divergence is utilized to compare the actual distribution and the estimated models via the \ac{EM} algorithm. \ac{KL}-divergence gets a higher value when two distributions have less similarities.
%%%%%%%%%%%%%%%%%%%%%%%%%%%%%%%% Error Metrics %%%%%%%%%%%%%%%%%%%%%%%%

%%%%%%%%%%%%%%%%%%%%%%%%%%%%%%%% MEASUREMENT SETUP %%%%%%%%%%%%%%%%%%%%%%%%
\section{Channel Measurement Campaign and Data Processing}\label{sec:measurement_setup}

The measurement setup exihibited in \FGR{fig_meas_env} is allocated in one of the anechoic chambers of Turkish Science Foundation \cite{miltal_lab} with the dimensions of  $7m\times4m\times3m$ to make sure that the \ac{LOS} components of the transmissions are observed and received properly.

\begin{figure}[!t]
	\centering
	\includegraphics[width=3.45in,height=2.05in]{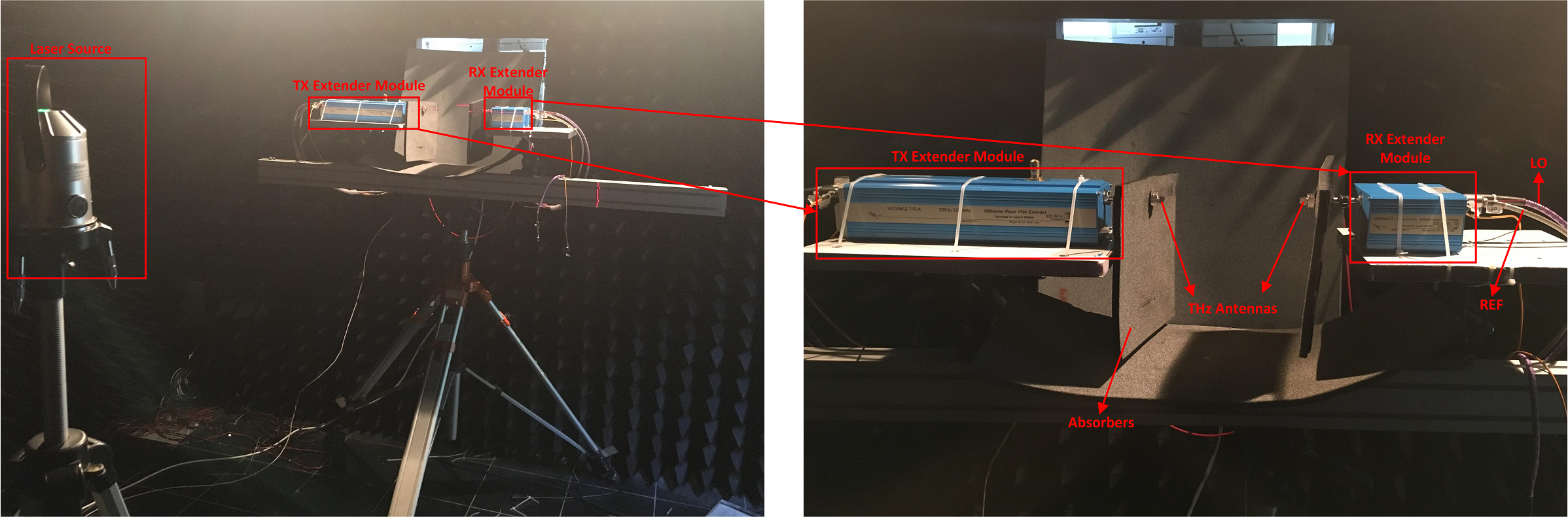}
	\caption{Measurement setup is prepared in the anechoic chamber to suppress possible reflections and guarantee \ac{LOS} conditions. Laser source is used to eliminate any misalignment.}
	\label{fig_meas_env}
\end{figure}

The setup is comprised of four main hardware components: \begin{inparaenum}[(i)]\item{a \ac{PNA} \ac{VNA}, which is coded as E$8361$A}, \item{extender modules for millimeter wave propagation, i.e., V$03$VNA$2$-A and V$03$VNA$2$-T/R-T coded devices from \ac{OML}}, and \item{N$5260$ coded controller for extenders from the same company}\end{inparaenum}. \ac{VNA} can analyze signals up to \unit{$67$}{GHz}, therefore, extender modules are utilized to be able to cover the \unit{$220$}{GHz} to \unit{$325$}{GHz} bands. The V$03$VNA$2$-T/R-A has 18 multipliers that can modulate signals on the \unit{$10$}{GHz} to \unit{$20$}{GHz} range up to the \unit{$300$}{GHz} region. On the contrary, the transmitted signal from the wireless channel is down-converted via V$03$VNA$2$-T by using the same number of mixers and the resultant signal is at the \ac{IF} of \unit{$5$}{MHz} to \unit{$300$}{MHz}. Following the down-conversion, the \ac{IF} signal is provided as input to the \ac{VNA}. The channel characteristics analysis is conducted considering the difference between the characteristics of transmitted and received signals. Please also note that we utilize a laser level tool to ensure that both transmitter and receiver are perfectly aligned for \ac{LOS} transmission. The block diagram of this process is depicted in \FGR{fig_sim}.

\begin{figure}[!t]
	\centering
	\includegraphics[width=3.45in,height=2.05in]{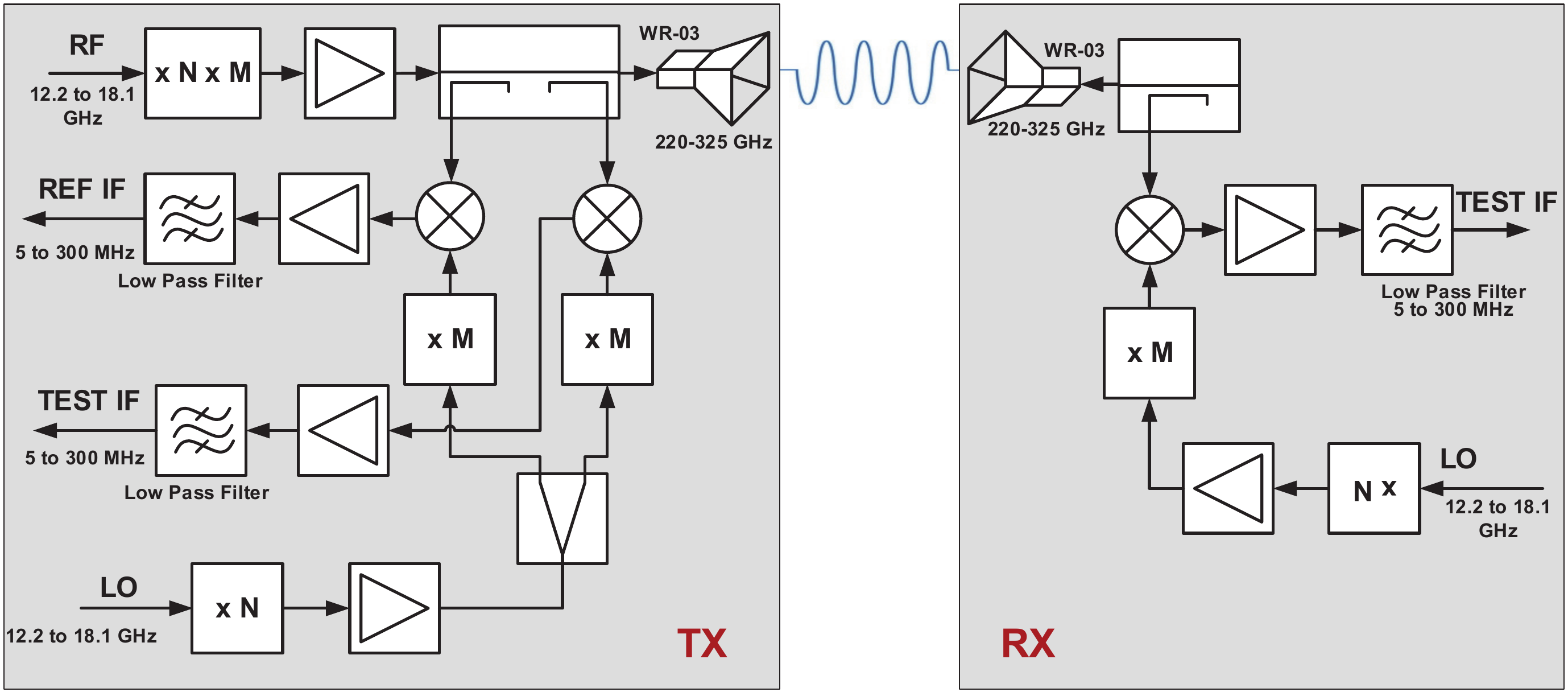} %width=3.35in,height=1.55in
	\caption{Block diagram for the measurement setup, which uses bottom-up approach to generate \ac{THz} signals.}
	\label{fig_sim}
\end{figure}

\begin{figure*}[ht]
    \centering
    \subfigure[]{%
        \label{fig:thz_20cm}%
        \includegraphics[width=0.31\textwidth]{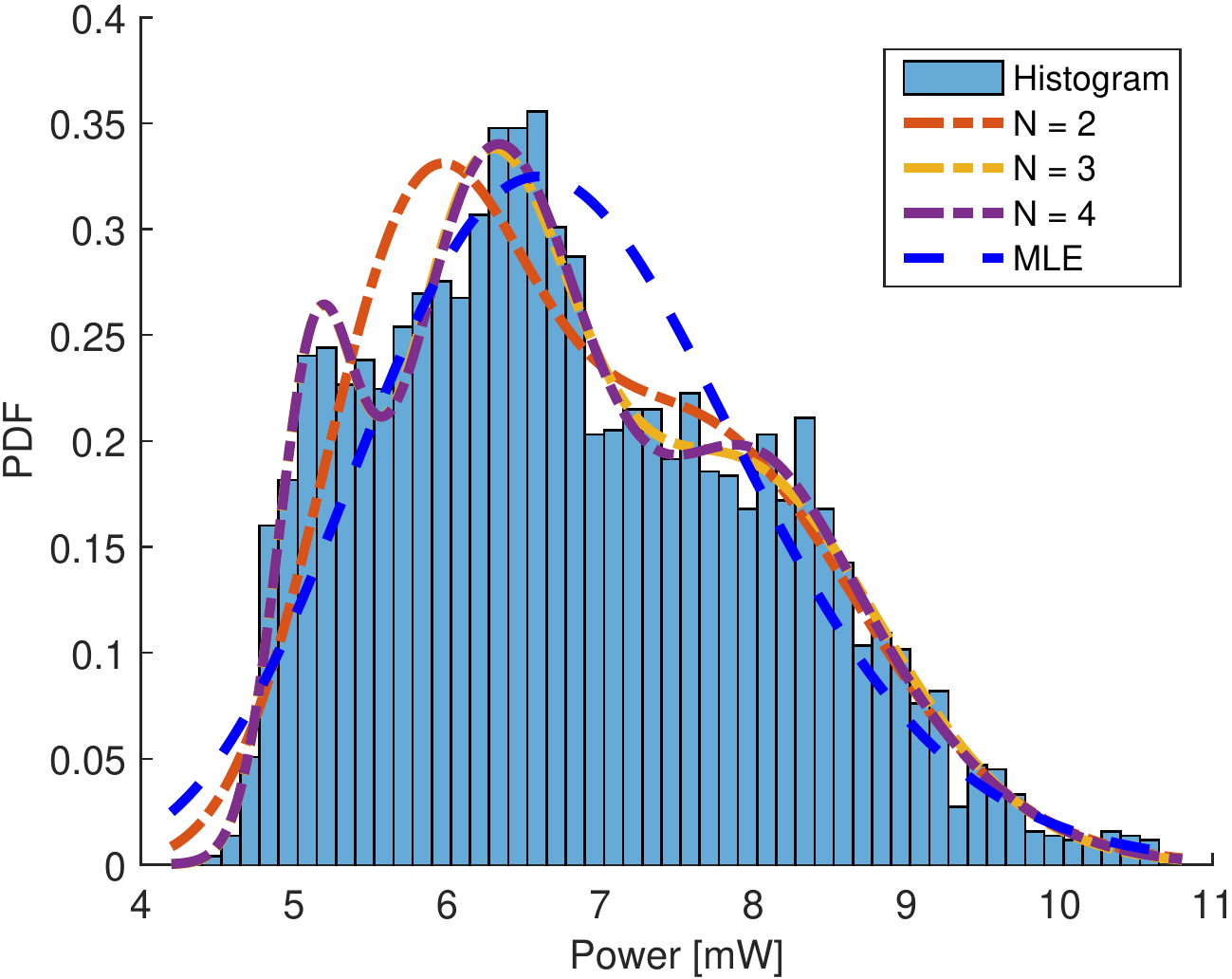}}
    \quad
    \subfigure[]{%
        \label{fig:thz_30cm}%
        \includegraphics[width=0.31\textwidth]{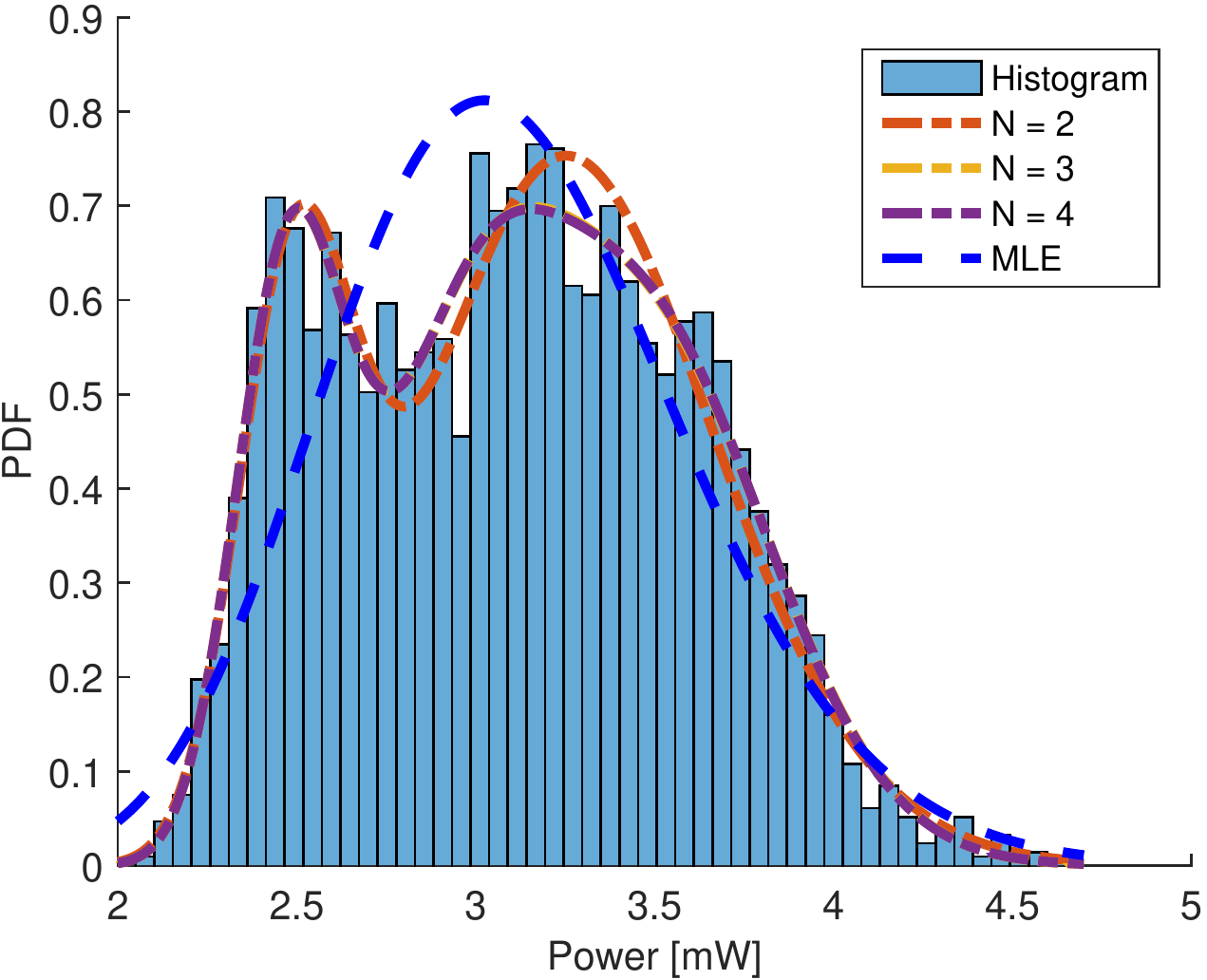}}%
    \quad
    \subfigure[]{%
        \label{fig:thz_40cm}%
        \includegraphics[width=0.31\textwidth]{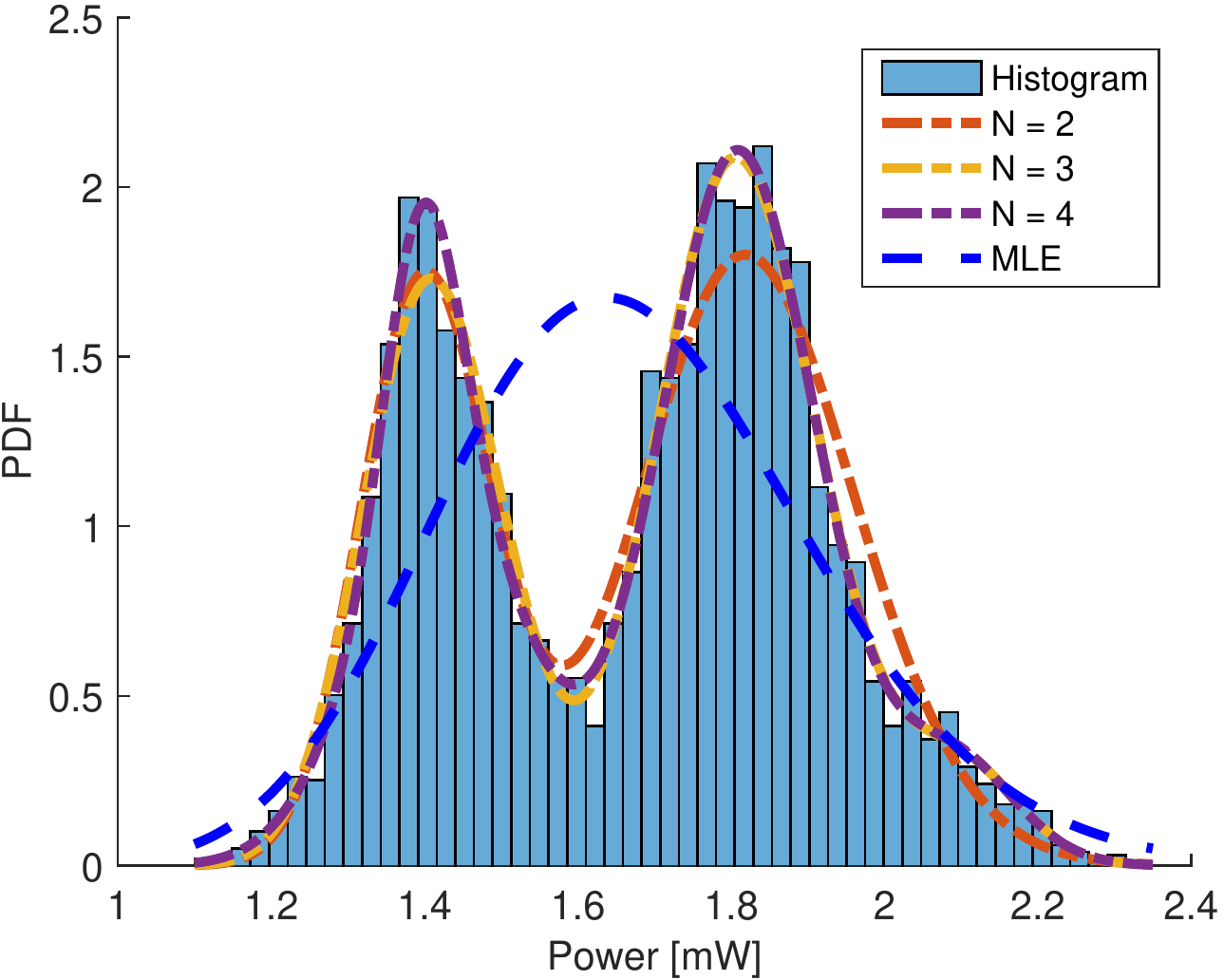}}%%
    \quad
    \subfigure[]{%
        \label{fig:thz_60cm}%
        \includegraphics[width=0.31\textwidth]{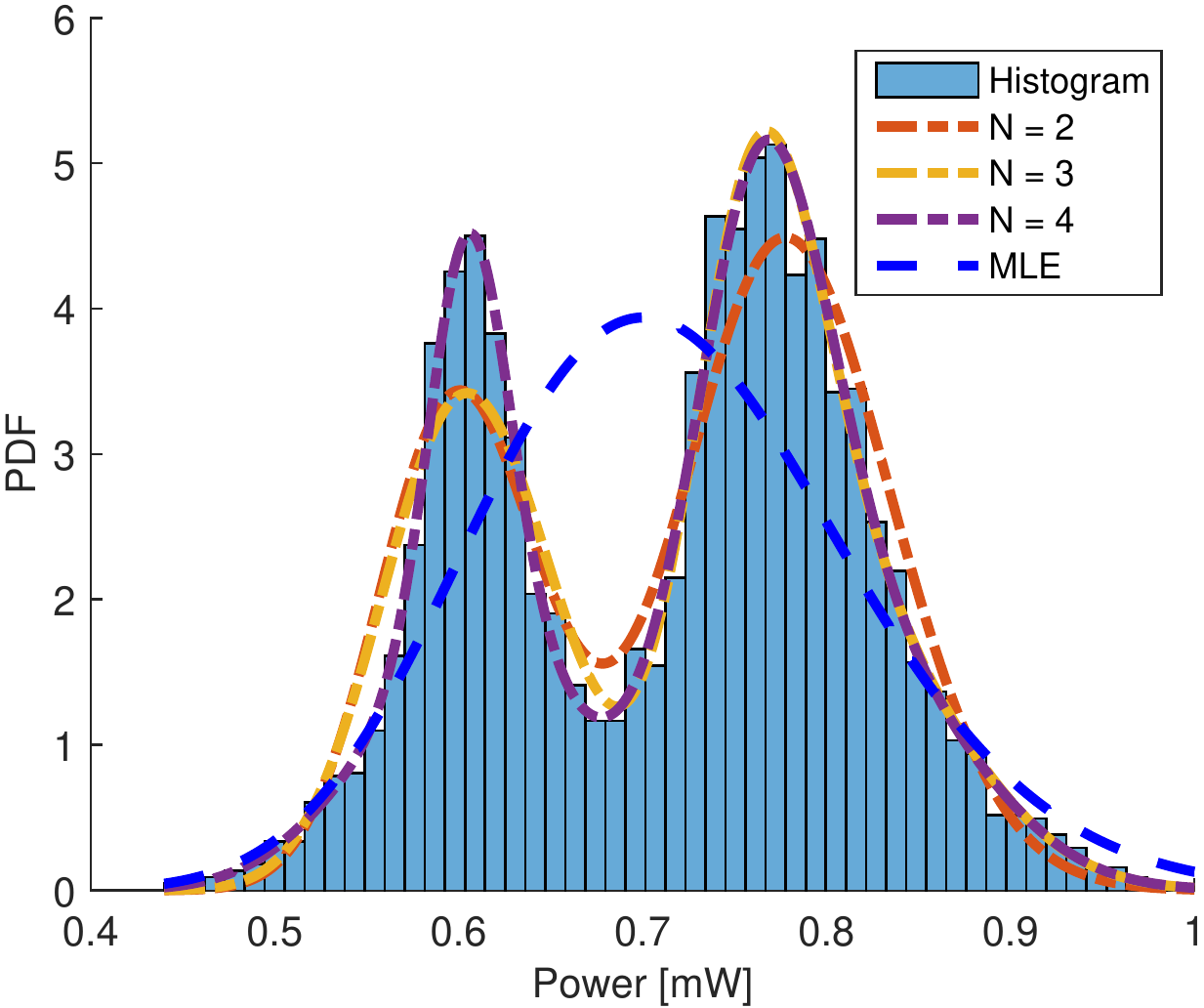}}%
    \quad
    \subfigure[]{%
        \label{fig:thz_80cm}%
        \includegraphics[width=0.31\textwidth]{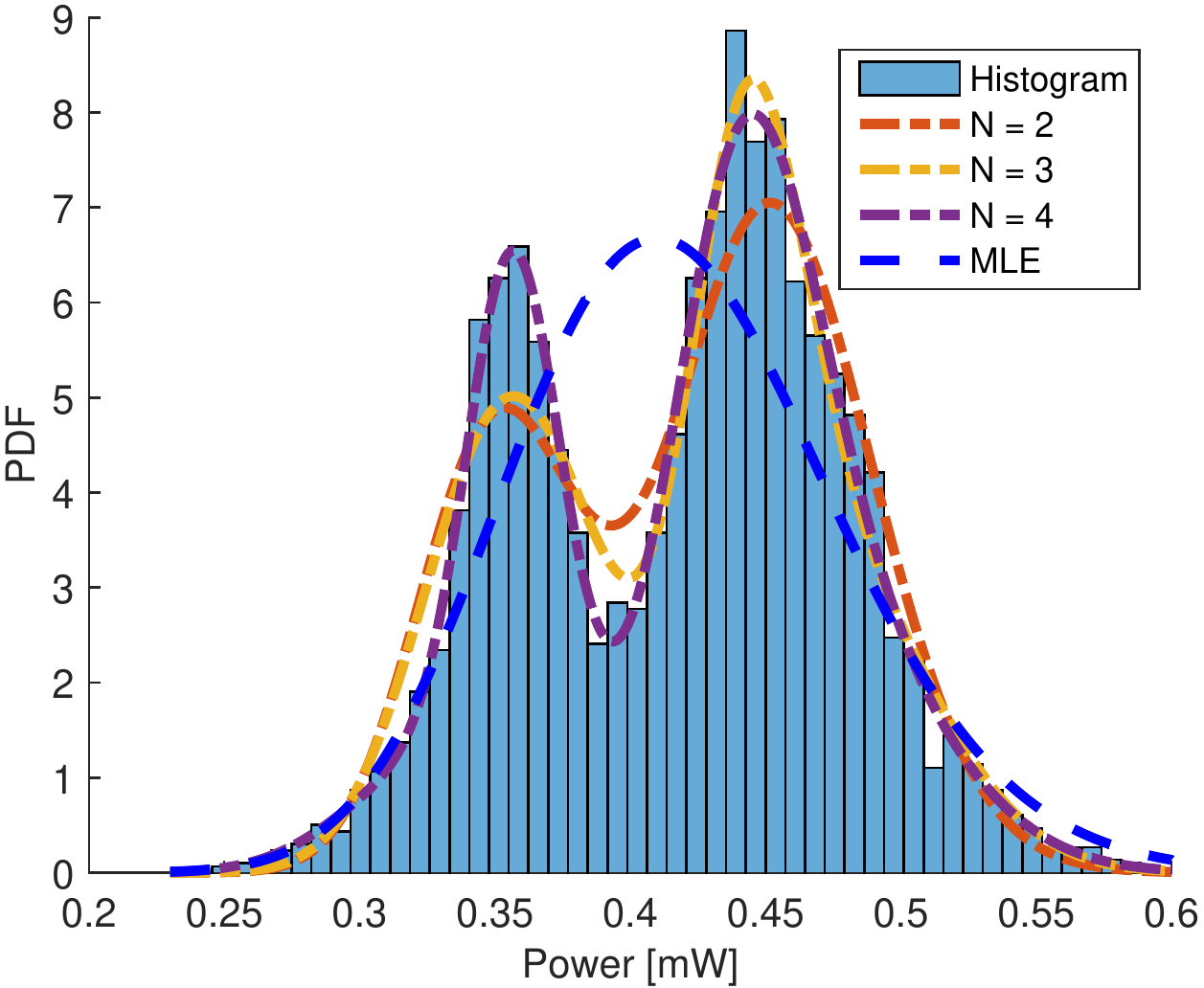}}%
    \caption{Mixture models fit much better to the measurement data for (a) \unit{$20$}{cm}, (b) \unit{$30$}{cm}, (c) \unit{$40$}{cm},  (d) \unit{$60$}{cm}, and (e) \unit{$80$}{cm} compared to \ac{MLE}.}
    \label{fig:mixtures}
\end{figure*}

When the hardware characteristics are considered, it is seen that the typical source match at the output is \unit{$9$}{dB} for balanced multipliers, which are connected to the WR-$10$ band extension multiplier chains. Each chain contains WR-$03$ wave-guide output interfaces. The signal generated can be a \ac{CW} or frequency sweeping signal. The level of RF power for the \ac{LO} to run \ac{OML} modules should be in \unit{$+10$}{dBm} range. The extender characteristics are as follows; the phase stability is \unit{$\pm 0.4$}{dB} in the range of \unit{$\pm 8$}{\si{\degree}}, typical dynamic range is \unit{$75$}{dB} with the minimum of \unit{$60$}{dB}.  

At the earlier stages of the measurement campaign, we realized relatively small impairments in terms of phase and magnitude stability are observed at the range of \unit{$240$}{GHz} to \unit{$300$}{GHz} frequencies. Thus, we decided to utilize these bands to be able to achieve the best results for our purpose in this study. \acp{spar} are utilized to understand the channel transfer function of these bands and for the modelling of the wireless channel, first a calibration procedure is executed. In this context, a direct connection is established between the transmitting and receiving wave-guide ports of the extenders. Following this step, calibration data is saved inside the measurement devices of the setup and it is converted to the form of complex $S_{21}$ parameters of each point of measurement. The measurement system also includes cables and connectors whic are also separately calibrated to eliminate the impairments.

Two identical horn antennas with \unit{$24.8$}{dBi} gain at their center frequencies are connected at both transmitting and receiving ends of the measurement system. Therefore, this setup covered \unit{$60$}{GHz} band between \unit{$240$}{GHz} to \unit{$300$}{GHz} and recordings are done over $4096$ points utilizing an \ac{IF} \ac{BW} of \unit{$100$}{Hz}. Such process led to the improvement of observed dynamic range and reduction of the noise floor. Eventually \unit{$14.648$}{MHz} became the spectrum resolution available. Each set of captured \ac{I/Q} samples are transferred into a laptop computer. Necessary conversions are applied and all the analyses are done on MATLAB R2018b software to carry out the baseband operations for each transmitter-receiver separation distance given in \TAB{tab:metrics}.
%%%%%%%%%%%%%%%%%%%%%%%%%%%%%%%% MEASUREMENT SETUP %%%%%%%%%%%%%%%%%%%%%%%%

%%%%%%%%%%%%%%%%%%%%%%% GAMMA MIXTURE MODELS FOR THZ CH. %%%%%%%%%%%%%%%%%%

\section{Gamma Mixture Model for Terahertz Wireless Channels}\label{sec:gamma_mix_thz}

%%%%%%%%%%%%%%%%%%%%%%%%%%%%%%%%%%%%TABLE%%%%%%%%%%%%%%%%%%%%%%
\begin{table*}[]
\centering
\renewcommand{\arraystretch}{0.80}
	\caption{Error metrics for PDF estimations at distinct distances.} 
	\label{tab:metrics}
\begin{tabular}{cccccccc}
\toprule
\multirow{2}{*}{\textbf{Distance}}  & \multirow{2}{*}{\textbf{Mixture}} & \multicolumn{3}{c}{\textbf{Parameters}} & \multirow{2}{*}{\shortstack{\textbf{WMRD}\\($\times10^{-2}$)}}  & \multirow{2}{*}{\textbf{KL Divergence}} & \multirow{2}{*}{\shortstack{\textbf{KS Test}\\($p=0.05$)}} \\
                           &                           & $\pi$   & $\alpha$  & $\beta$  &                        &                                &                          \\ \midrule
\multirow{10}{*}{\unit{$20$}{cm}}    & MLE                      & 1.00    & 30.084    & 0.227    & 1.580                  & 4.635                          & \textit{Passed}                   \\ \cmidrule{2-8} 
                           & \multirow{2}{*}{N = 2}   & 0.540   & 72.285    & 0.0824   & \multirow{2}{*}{1.573} & \multirow{2}{*}{0.651}         & \multirow{2}{*}{\textit{Passed}}  \\
                           &                          & 0.460   & 67.904    & 0.115    &                        &                                &                          \\ \cmidrule{2-8} 
                           & \multirow{3}{*}{N = 3}   & 0.463   & 116.797   & 0.0539   & \multirow{3}{*}{1.572} & \multirow{3}{*}{0.813}         & \multirow{3}{*}{\textit{Passed}}  \\
                           &                          & 0.397   & 85.985    & 0.093    &                        &                                &                          \\
                           &                          & 0.140   & 406.051   & 0.012    &                        &                                &                          \\ \cmidrule{2-8} 
                           & \multirow{4}{*}{N = 4}   & 0.538   & 100.310   & 0.063    & \multirow{4}{*}{1.572} & \multirow{4}{*}{0.797}         & \multirow{4}{*}{\textit{Passed}}  \\
                           &                          & 0.193   & 185.751   & 0.042    &                        &                                &                          \\
                           &                          & 0.137   & 407.257   & 0.012    &                        &                                &                          \\
                           &                          & 0.132   & 120.786   & 0.072    &                        &                                &                          \\ \midrule
\multirow{10}{*}{\unit{$30$}{cm}}    & MLE                      & 1.00    & 39.060    & 0.079    & 1.541                  & 3.881                          & \textit{Passed}                   \\ \cmidrule{2-8} 
                           & \multirow{2}{*}{N = 2}   & 0.752   & 67.765    & 0.048    & \multirow{2}{*}{1.536} & \multirow{2}{*}{0.822}         & \multirow{2}{*}{\textit{Passed}}  \\
                           &                          & 0.248   & 236.829   & 0.010    &                        &                                &                          \\ \cmidrule{2-8} 
                           & \multirow{3}{*}{N = 3}   & 0.388   & 123.224   & 0.024    & \multirow{3}{*}{1.536} & \multirow{3}{*}{0.906}         & \multirow{3}{*}{\textit{Passed}}  \\
                           &                          & 0.370   & 132.766   & 0.026    &                        &                                &                          \\
                           &                          & 0.242   & 259.535   & 0.009    &                        &                                &                          \\ \cmidrule{2-8} 
                           & \multirow{4}{*}{N = 4}   & 0.303   & 144.100   & 0.020    & \multirow{4}{*}{1.536} & \multirow{4}{*}{0.903}         & \multirow{4}{*}{\textit{Passed}}  \\
                           &                          & 0.250   & 254.633   & 0.009    &                        &                                &                          \\
                           &                          & 0.234   & 104.663   & 0.032    &                        &                                &                          \\
                           &                          & 0.213   & 139.345   & 0.025    &                        &                                &                          \\ \midrule
\multirow{10}{*}{\unit{$40$}{cm}}    & MLE                      & 1.00    & 49.334    & 0.034    & 1.497                  & 3.349                          & \textit{Passed}                   \\ \cmidrule{2-8} 
                           & \multirow{2}{*}{N = 2}   & 0.626   & 172.946   & 0.010    & \multirow{2}{*}{1.486} & \multirow{2}{*}{1.118}         & \multirow{2}{*}{\textit{Passed}}  \\
                           &                          & 0.374   & 269.180   & 0.005    &                        &                                &                          \\ \cmidrule{2-8} 
                           & \multirow{3}{*}{N = 3}   & 0.550   & 295.648   & 0.006    & \multirow{3}{*}{1.484} & \multirow{3}{*}{1.067}         & \multirow{3}{*}{\textit{Passed}}  \\
                           &                          & 0.396   & 240.228   & 0.006    &                        &                                &                          \\
                           &                          & 0.054   & 767.221   & 0.002    &                        &                                &                          \\ \cmidrule{2-8} 
                           & \multirow{4}{*}{N = 4}   & 0.532   & 322.175   & 0.005    & \multirow{4}{*}{1.483} & \multirow{4}{*}{1.016}         & \multirow{4}{*}{\textit{Passed}}  \\
                           &                          & 0.290   & 160.954   & 0.009    &                        &                                &                          \\
                           &                          & 0.119   & 782.971   & 0.002    &                        &                                &                          \\
                           &                          & 0.059   & 720.197   & 0.003    &                        &                                &                          \\ \midrule
\multirow{10}{*}{\unit{$60$}{cm}}    & MLE                      & 1.00    & 49.026    & 0.014    & 1.446                  & 2.646                          & \textit{Passed}                   \\ \cmidrule{2-8} 
                           & \multirow{2}{*}{N = 2}   & 0.626   & 196.498   & 0.004    & \multirow{2}{*}{1.435} & \multirow{2}{*}{1.351}         & \multirow{2}{*}{\textit{Passed}}  \\
                           &                          & 0.374   & 191.389   & 0.003    &                        &                                &                          \\ \cmidrule{2-8} 
                           & \multirow{3}{*}{N = 3}   & 0.340   & 449.070   & 0.002    & \multirow{3}{*}{1.432} & \multirow{3}{*}{1.226}         & \multirow{3}{*}{\textit{Passed}}  \\
                           &                          & 0.399   & 170.063   & 0.035    &                        &                                &                          \\
                           &                          & 0.201   & 230.398   & 0.036    &                        &                                &                          \\ \cmidrule{2-8} 
                           & \multirow{4}{*}{N = 4}   & 0.405   & 163.658   & 0.004    & \multirow{4}{*}{1.429} & \multirow{4}{*}{1.213}         & \multirow{4}{*}{\textit{Passed}}  \\
                           &                          & 0.356   & 428.192   & 0.002    &                        &                                &                          \\
                           &                          & 0.189   & 839.961   & 0.001    &                        &                                &                          \\
                           &                          & 0.050   & 541.486   & 0.0016   &                        &                                &                          \\ \midrule
\multirow{10}{*}{\unit{$80$}{cm}}    & MLE                      & 1.00    & 48.192    & 0.008    & 1.392                  & 2.227                          & \textit{Passed}                   \\ \cmidrule{2-8} 
                           & \multirow{2}{*}{N = 2}   & 0.634   & 158.089   & 0.003    & \multirow{2}{*}{1.388} & \multirow{2}{*}{1.725}         & \multirow{2}{*}{\textit{Passed}}  \\
                           &                          & 0.366   & 135.815   & 0.002    &                        &                                &                          \\ \cmidrule{2-8} 
                           & \multirow{3}{*}{N = 3}   & 0.412   & 119.250   & 0.003    & \multirow{3}{*}{1.384} & \multirow{3}{*}{1.590}         & \multirow{3}{*}{\textit{Passed}}  \\
                           &                          & 0.316   & 462.180   & 0.001    &                        &                                &                          \\
                           &                          & 0.272   & 180.013   & 0.0027   &                        &                                &                          \\ \cmidrule{2-8} 
                           & \multirow{4}{*}{N =4}    & 0.432   & 281.987   & 0.0016   & \multirow{4}{*}{1.381} & \multirow{4}{*}{1.331}         & \multirow{4}{*}{\textit{Passed}}  \\
                           &                          & 0.222   & 441.924   & 0.0008   &                        &                                &                          \\
                           &                          & 0.204   & 161.406   & 0.003    &                        &                                &                          \\
                           &                          & 0.142   & 76.095    & 0.0045   &                        &                                &                          \\  \bottomrule
\end{tabular}
\end{table*}
%%%%%%%%%%%%%%%%%%%%%%%%%%%%%%%%%%%%TABLE%%%%%%%%%%%%%%%%%%%%%%

In this section, Gamma mixture models are employed to model received power distribution for five measurement described in Section~\ref{sec:measurement_setup}. The received power, $P_{rx}$, is calculated in the linear scale as
\begin{align}
    P_{rx} = |S_{21}|^{2}P_{tx},
\end{align}
where $P_{tx}$ is the transmitted signal power and constant during the transmission time. $|S_{21}|$ denotes the amplitude response of the propagation channel. It is known that the instantaneous \ac{SNR} for a signal with bandwidth of $W$ is defined as:
\begin{align}
    \gamma = \frac{P_{rx}}{WN_{0}}
\end{align}
under the \ac{AWGN} with power spectral density $N_{0}/2$. Therefore, \ac{SNR} is related to the fading channel parameters, as well as the received power. By utilizing the instantaneous \ac{SNR}, it is possible to derive the channel outage probability and the channel capacity \cite{goldsmith2005wireless}.

\begin{figure*}[!t]
    \centering
    \subfigure[]{%
        \label{fig:thz_20cm_comp}%
        \includegraphics[width=0.3\textwidth]{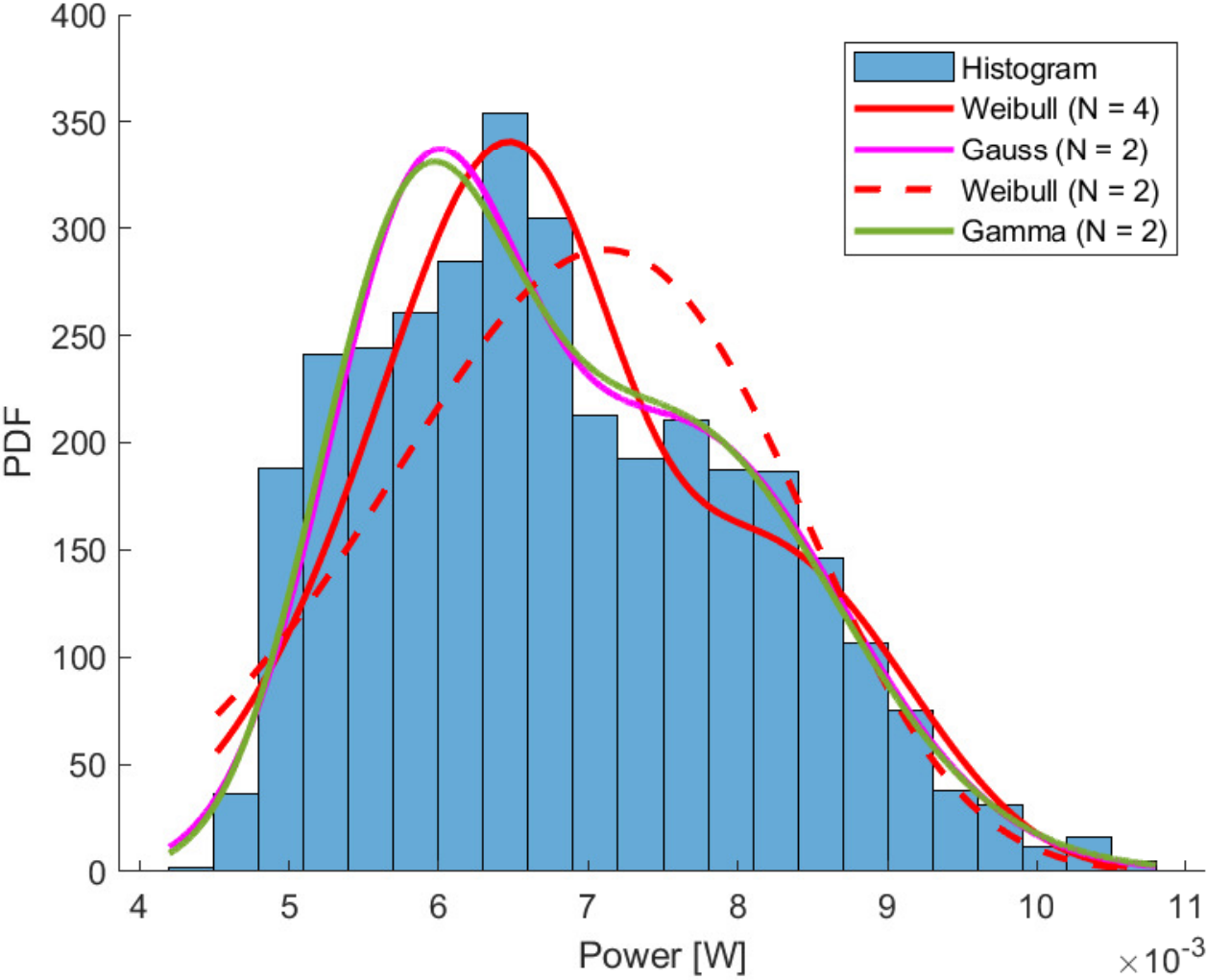}}
    \qquad
    \subfigure[]{%
        \label{fig:thz_60cm_comp}%
        \includegraphics[width=0.3\textwidth]{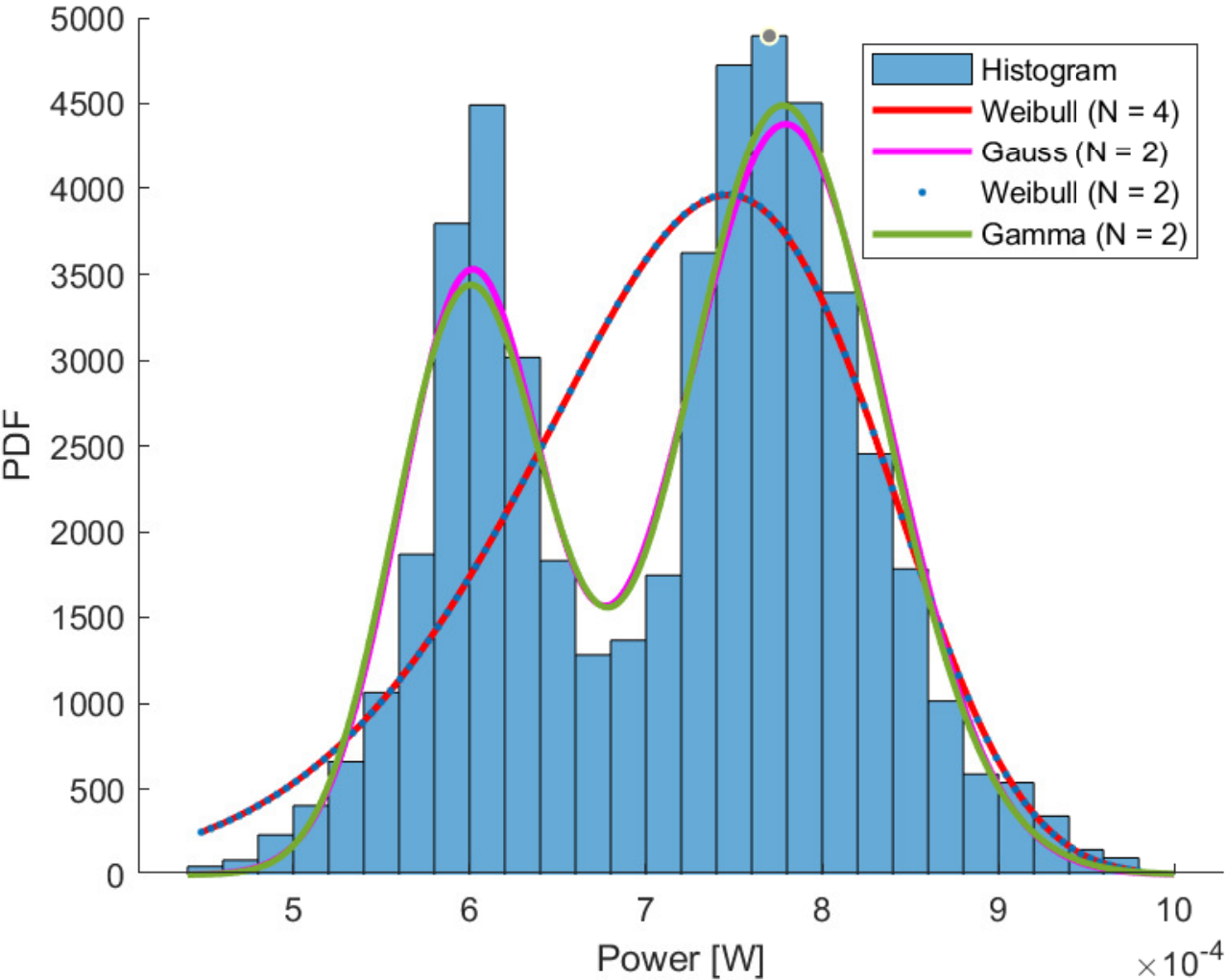}}%
    \qquad
    \subfigure[]{%
        \label{fig:thz_all_comp}%
        \includegraphics[width=0.3\textwidth]{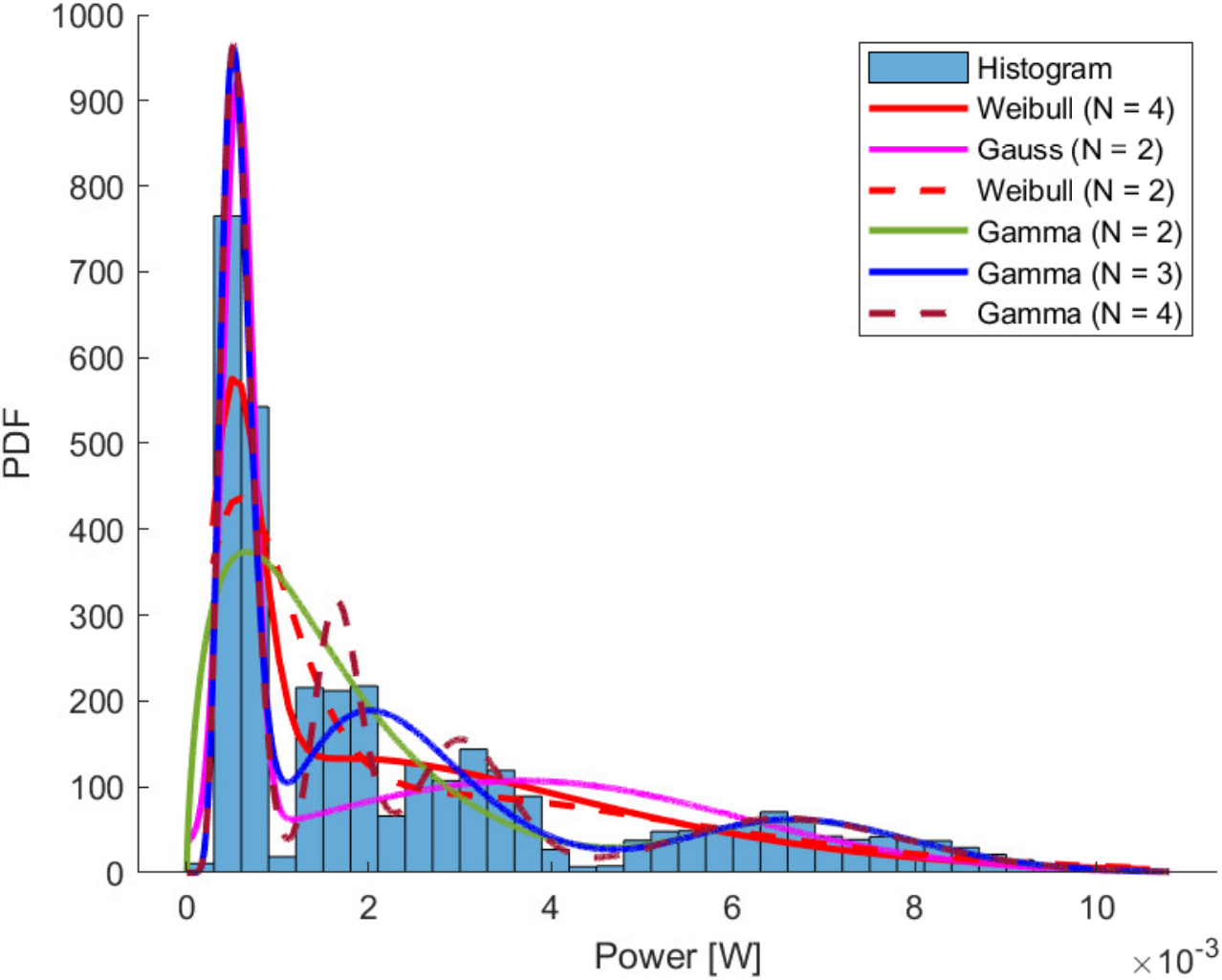}}
    \caption{Gamma mixtures fit much better to the measurement data for (a) \unit{$20$}{cm}, (b) \unit{$60$}{cm}, (c) all measurement data compared to Gaussian and Weibull mixtures.}
    \label{fig:mixtures_comparison}
\end{figure*}

\begin{table}[!t]
\centering
	\caption{KL Divergence results for Gamma, Gaussian and Weibull mixtures.}
	\label{tab:comparison}
\begin{tabular}{ccc}
\toprule
\textbf{Distance}      & \textbf{Mixture} & \textbf{KL Divergence} \\ \midrule
\multirow{4}{*}{$20$cm} & Gaussian (N = 2)    & 0.655                  \\ \cmidrule{2-3} 
                       & Weibull (N = 2)  & 1.726                  \\ \cmidrule{2-3} 
                       & Weibull (N = 4)  & 0.978                  \\ \cmidrule{2-3} 
                       & Gamma (N = 2)    & 0.651                  \\ \midrule
\multirow{4}{*}{$60$cm} & Gaussian (N = 2)    & 1.344                  \\ \cmidrule{2-3} 
                       & Weibull (N = 2)  & 2.321                  \\ \cmidrule{2-3} 
                       & Weibull (N = 4)  & 2.321                  \\ \cmidrule{2-3}
                       & Gamma (N = 2)    & 1.351                  \\ \midrule
\multirow{6}{*}{All}   & Gaussian (N = 2)    & 0.582                  \\ \cmidrule{2-3} 
                       & Weibull (N = 2)  & 1.436                  \\ \cmidrule{2-3} 
                       & Weibull (N = 4)  & 1.427                  \\ \cmidrule{2-3} 
                       & Gamma (N = 2)    & 1.023                  \\ \cmidrule{2-3} 
                       & Gamma (N = 3)    & 0.564                  \\ \cmidrule{2-3} 
                       & Gamma (N = 4)    & 0.504                  \\ \bottomrule
\end{tabular}
\end{table}

\subsection{Gamma Mixture Model Results}\label{sec:gamma_mixture_results}

In order to model the received signal power, both \ac{MLE} and the \ac{EM} algorithm are used. \ac{EM} algorithm enables to determine the parameters of the estimated mixture components for the measurements. Based on the aforementioned points stated in \SEC{sec:gamma_why}, the Gamma distributions are utilized because of the facts that its \ac{MGF} is tractable and there is an approximation for small-scale fading channels. 

In \FGR{fig:mixtures}, it can be clearly seen that \ac{MLE} estimation is not a good fit for the measured histogram; however, the mixture models fit better. For example, it can be said that three mixtures of Gamma distribution are sufficient to estimate the actual histogram for the distance of \unit{$20$}{cm}. However, \ac{MLE} gets hampered to fit since it assumes that there is no serious change in the channel behavior through the transmission band due to the characteristics of the molecules in the propagation environment. Even though this assumption can be reasonable for the cellular communication bands below \unit{$60$}{GHz}, it is not held for the \ac{THz} band. As shown in \FGR{fig:mixtures}, the mixture models provide more adequate \acp{PDF} than \ac{MLE} for the tx-rx separation between \unit{$20$}{cm} and \unit{$80$}{cm}. \FGR{fig:mixtures} shows that three mixtures and four mixtures of Gamma distribution have almost the same performance to fit the actual histograms. It is seen in \FGR{fig:mixtures} that the mean of the received signal power decreases for increasing distance, as expected. Furthermore, \FGR{fig:mixtures} clearly exhibits different clusters in the histograms especially for the distances longer than \unit{$30$}{cm}, which is consistent with \cite{akyildiz2014terahertz, tekbiyik2019terahertz}.

Moreover, \ac{WMRD} results and \ac{KL}-divergence also confirm that mixture models converge to the actual histogram better than \acp{MLE}. \ac{WMRD} results, \ac{KL} distance, and \ac{KS} test results are presented in \TAB{tab:metrics}. \ac{WMRD} results are not accurate enough to show the difference between \ac{MLE} and mixture models. \ac{WMRD} can demonstrate that the mixture model is more successful than \ac{MLE} with only a small variation in its value. However, \ac{KL}-divergence creates metrics more sensitive to differences between mixture models and \acp{MLE}. For instance, \ac{KL}-divergence is found as $4.635$ for \ac{MLE} at \unit{$20$}{cm}, whereas it is $0.651$ for two mixtures. \ac{KL}-divergences show that the models consisting four mixtures are more similar to the actual \acp{PDF} for all measurements except \unit{$20$}{cm}. Surprisingly, \ac{KL}-divergence of the model with two mixtures is the smallest for the measurement with the distance of \unit{$20$}{cm}. Furthermore, the results obtained from goodness-of-fit
test with the confidence level $p = 0.05$ imply the suitability of the mixture models to actual \acp{PDF}.

\subsection{Comparison with Weibull and Gaussian Distributions}

Although it is shown that Gamma mixture distribution is able to describe THz channels with ultra broadband in the \SEC{sec:gamma_mixture_results}, we instigate the accuracy of THz channel models with various mixture of distributions rather than Gamma distribution. We evaluate the mixtures of normal and Weibull distributions. \ac{EM} simulations have been performed for three measurement data.

Firstly, $20$cm measurement data is evaluated. It is observed in~\FGR{fig:thz_20cm_comp} that the mixture of two Gamma distributions fits the measurement better. On the other hand, Weibull mixture with two distinct distributions cannot represent the actual data; however, increasing the number of mixtures improves accuracy as seen in~\FGR{fig:thz_20cm_comp}. Gaussian mixtures are observed to describe the channel almost as well as Gamma mixture. However, Gamma mixture has a slightly smaller \ac{KL} divergence.

$60$cm measurements are analyzed with the same scenario as the former. In these measurements, Weibull mixtures are unable to provide adequate fitting for modeling the channel. Surprisingly, increasing the number of mixtures could not improve the accuracy of the channel model as depicted in~\FGR{fig:thz_60cm_comp}. Particularly, Gaussian mixture slightly outperforms the accuracy of Gamma mixture modeling in terms of \ac{KL} divergence. 

Finally, \ac{EM} algorithm for all measurement data in~\FGR{fig:mixtures_comparison} are evaluated. When two distinct distributions are employed, Gaussian mixture shows the best fitting performance. But, it is worth noting the \ac{CDF} of Gaussian distribution cannot be evaluated as in a closed form while Gamma distribution enables closed form expressions. On the other hand, Weibull distribution cannot leverage fitting to actual data even if four mixtures are employed. It is shown that by increasing the number of Gamma distributions in mixture, fitting to measurements can be done in a more accurate way. \ac{KL} divergence values between fitted mixtures and measurements are summarized in~\TAB{tab:comparison}.

%%%%%%%%%%%%%%%%%%%%%%%%%%%%%%%% CONCLUSION %%%%%%%%%%%%%%%%%%%%%%%%
\section{Concluding Remarks and Future Works}\label{sec:conclusion}
In this paper, we investigate the channel model for the \ac{THz} band in between \unit{$240$}{GHz}-\unit{$300$}{GHz} by using Gamma mixture models. To find the mixture parameters, \ac{EM} algorithm is utilized. It is visible that the mixture models are better to fit the measurement histogram for all measurements compared to \acp{MLE}. The comparison between the mixture models and the actual \acp{PDF} is carried out by \ac{WMRD}, \ac{KS} and \ac{KL}-divergence metrics. The metrics adminiculate that the mixture of Gamma distributions can accurately model the \ac{THz} channels.

Since the average channel capacity, the outage probability, and the symbol error rate are derived for mixture Gamma wireless channels, the analytical analyse can be carried by using mixture parameters given in this study. As known, the \ac{EM} algorithm requires the number of mixtures as a priori information. However, to determine the number of mixtures, the Dirichlet process mixture model and Bayesian information criterion can be utilized. 

Due to limitations of the measurement setup in this study, any mobility could not be considered. However, measurement campaigns which enable to consider mobility in THz band should be carried out. Furthermore, in-vivo channel characteristics are heavily dependent on the density of the materials in the tissue; therefore, the in-vivo channels have greater changes in their behaviors. It can be thought that mixture models are appropriate for also in-vivo channels. As a future work, in-vivo channel can be investigated by utilizing mixture models.
%%%%%%%%%%%%%%%%%%%%%%%%%%%%%%%% CONCLUSION %%%%%%%%%%%%%%%%%%%%%%%%
%%%%%%%%%%%%%%%%% REFERENCES %%%%%%%%%%%%%%%%%%%%%%

\bibliographystyle{IEEEtran}
\bibliography{tvt_terahertz}

%%%%%%%%%%%%%%%}%% BIOGRAPHIES %%%%%%%%%%%%%%%%%%%%%%

\vspace{-0.5in}
\begin{IEEEbiography}
    [{\includegraphics[width=1in,height=1.25in,clip,keepaspectratio]{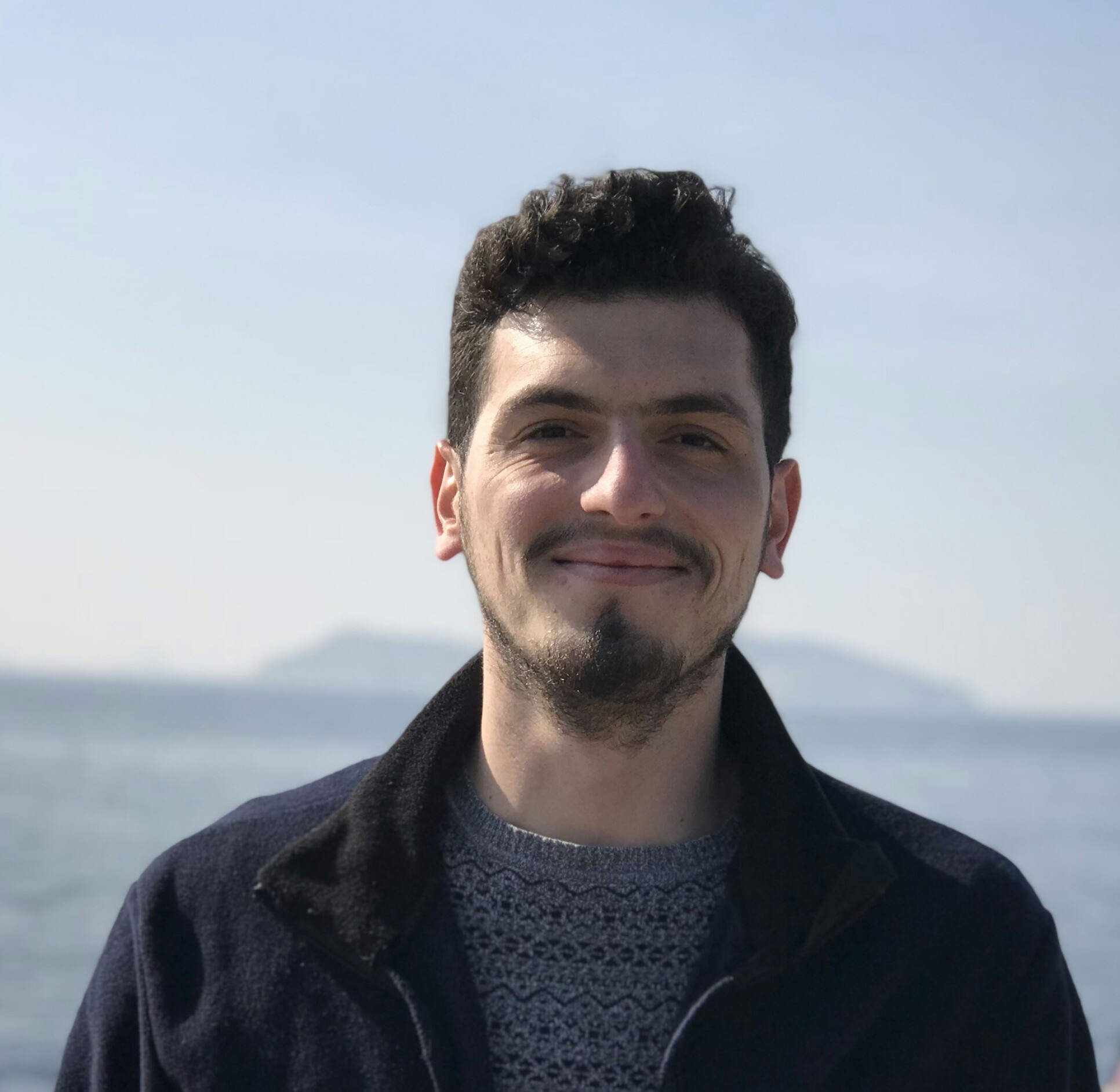}}]{K{\"{u}}r{\c{s}}at Tekb{\i}y{\i}k} [StM'19] (tekbiyik@itu.edu.tr) received his B.Sc. and M.Sc. degrees (with high honors) in electronics and communication engineering from Istanbul Technical University, Istanbul, Turkey, in 2017 and 2019, respectively. He is currently pursuing his Ph.D. degree in telecommunications engineering in Istanbul Technical University. He is also researcher at TUBITAK BILGEM. His research interests are next-generation wireless communication systems, terahertz wireless communications, and machine learning.
\end{IEEEbiography}

\begin{IEEEbiography}
    [{\includegraphics[width=1in,height=1.25in,clip,keepaspectratio]{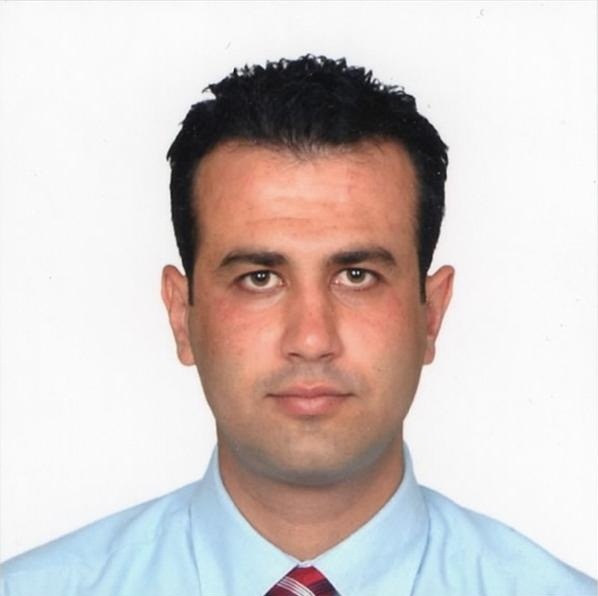}}]{Ali R{\i}za Ekti} is from Tarsus, Turkey. He received B.Sc. degree in Electrical and Electronics Engineering from Mersin University, Mersin, Turkey, (September 2002-June 2006), also studied at Universidad Politechnica de Valencia, Valencia, Spain in 2004-2005, received M.Sc. degree in Electrical Engineering from the University of South Florida, Tampa, Florida (August 2008-December 2009) and received Ph.D. degree in Electrical Engineering from Texas A\&M University (August 2010-August 2015). He is currently an assistant professor at Balikesir University Electrical and Electronics Engineering Department and also senior researcher at TUBITAK BILGEM. His current research interests include statistical signal processing, convex optimization, machine learning, resource allocation and traffic offloading in wireless communications in 4G and 5G systems and smart grid design and optimization.
\end{IEEEbiography}

\begin{IEEEbiography}
    [{\includegraphics[width=1in,height=1.25in,clip,keepaspectratio]{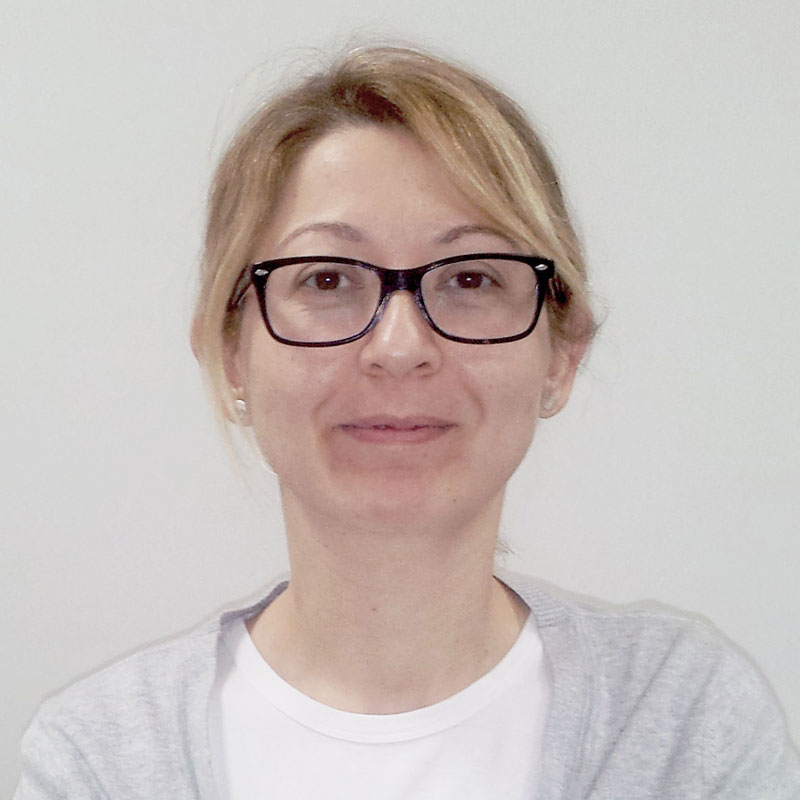}}]{G{\"{u}}ne{\c{s}} Karabulut Kurt} received a Ph.D. degree in electrical engineering from the University of Ottawa, Ottawa, ON, Canada, in 2006. Between 2005 and 2008, she was with TenXc Wireless, and Edgewater Computer Systems, in Ottawa Canada. From 2008 to 2010, she was with Turkcell R\&D, Istanbul. Since 2010, she has been with Istanbul Technical University. She is also an Adjunct Research Professor at Carleton University. She is serving as an Associate Technical Editor of \textit{IEEE Communications Magazine}.  
\end{IEEEbiography}
 
\begin{IEEEbiographynophoto}{Ali G\"{o}r\c{c}in} graduated from Istanbul Technical University with B.Sc. in Electronics and Telecommunications Engineering and completed his master's degree on defense technologies at the same university. After working at Turkish Science Foundation (TUBITAK) on avionics projects for more than six years, he moved to the U.S. to pursue PhD degree in University of South Florida (USF) on wireless communications. He worked for Anritsu Company during his tenure in USF and worked for Reverb Networks and Viavi Solutions after his graduation. He is currently holding an assistant professorship position at Yildiz Technical University in Istanbul and also serving as the president of Informatics and Information Security Research Center (BILGEM) of TUBITAK, responsible from testing and evaluation along with research and development activities on wireless communications technologies.

\end{IEEEbiographynophoto}

\begin{IEEEbiography}
    [{\includegraphics[width=1in,height=1.25in,clip,keepaspectratio]{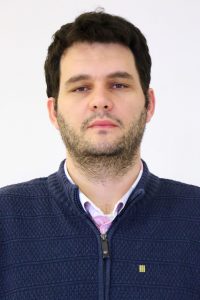}}]{Serhan Yarkan} received the B.S. and M.Sc. degrees in computer science from Istanbul University, Istanbul, Turkey, in 2001 and 2003, respectively, and the Ph.D. degree from the University of South Florida, Tampa, FL, USA, in 2009. He was a Post-Doctoral Research Associate with the Department of Computer and Electrical Engineering, Texas A\&M University, College Station, TX, USA, from 2010 to 2012. He is currently an Associate Professor with the Department of Electrical-Electronics Engineering, Istanbul Commerce University, Istanbul. His current research interests include statistical signal processing, cognitive radio, wireless propagation channel measurement and modeling, cross-layer adaptation and optimization, and interference management in next-generation wireless networks and underground mine channels and disaster communications. 
\end{IEEEbiography}

\balance
\end{document}